\documentclass[twocolumn]{aastex63}
\usepackage{enumerate}
\usepackage{enumitem}
\shorttitle{Outflow Bubbles from Compact Binary Mergers Embedded in AGN}
\shortauthors{Kimura, Murase, Bartos}

\begin{document}
\title{Outflow Bubbles from Compact Binary Mergers Embedded in Active Galactic Nuclei:\\
Cavity Formation and the Impact on Electromagnetic Counterparts}

\author[0000-0003-2579-7266]{Shigeo S. Kimura}
\affiliation{Frontier Research Institute for Interdisciplinary Sciences, Tohoku University, Sendai 980-8578, Japan}
\affiliation{Astronomical Institute, Graduate School of Science, Tohoku University, Sendai 980-8578, Japan}

\author[0000-0002-5358-5642]{Kohta Murase}
\affiliation{Department of Physics, The Pennsylvania State University, University Park, Pennsylvania 16802, USA}
\affiliation{Department of Astronomy \& Astrophysics, The Pennsylvania State University, University Park, Pennsylvania 16802, USA}
\affiliation{Center for Multimessenger Astrophysics, Institute for Gravitation and the Cosmos, The Pennsylvania State University, University Park, Pennsylvania 16802, USA}
\affiliation{Center for Gravitational Physics, Yukawa Institute for Theoretical Physics, Kyoto, Kyoto 606-8502 Japan}
\author[0000-0001-5607-3637]{Imre Bartos}
\affiliation{Department of Physics, University of Florida, Geinsville, Florida, USA}




\begin{abstract}
We propose a novel scenario for possible electromagnetic (EM) emission by compact binary mergers in the accretion disks of active galactic nuclei (AGNs). 
Nuclear star clusters in AGNs are a plausible formation site of compact-stellar binaries (CSBs) whose coalescences can be detected through gravitational waves (GWs). We investigate the accretion onto and outflows from CSBs embedded in AGN disks. We show that these outflows are likely to create outflow ``cavities'' in the AGN disks before the binaries merge, which makes EM or neutrino counterparts much less common than would otherwise be expected. 
We discuss the necessary conditions for detectable EM counterparts to mergers inside the outflow cavities. If the merger remnant black hole experiences a high recoil velocity and can enter the AGN disk, it can accrete gas with a super-Eddington rate, newly forming a cavity-like structure. This bubble can break out of the disk within a day to a week after the merger. Such breakout emission can be bright enough to be detectable by current soft X-ray instruments, such as {\it Swift}-XRT and {\it Chandra}.
\end{abstract}



%
\section{Introduction}
LIGO \citep{2015CQGra..32g4001L} and Virgo \citep{2015CQGra..32b4001A} discovered over 30 binary black hole (BBH) mergers \citep{Abbott:2020gyp,Abbott:2020niy}, transforming our ability to study these cosmic events. Among these findings, three peculiar events stand out. GW190412 has a low mass ratio of $M_{\rm sec}/M_{\rm pri}\simeq0.25-0.31$, where $M_{\rm pri}$ and $M_{\rm sec}$ are the primary and secondary masses, respectively \citep{LIGOScientific:2020stg}. 
Another event, GW190814 \citep{Abbott:2020khf}, has a secondary mass $\sim2.6~M_\odot$ that is in the lower mass gap, which is difficult to explain with standard stellar evolution \citep[e.g.,][]{2011ApJ...741..103F,2012ApJ...757...55O}. The objects in GW190814 also have highly asymmetric masses, with $M_{\rm sec}/M_{\rm pri}\simeq0.11$. Finally, GW190521 \citep{Abbott:2020tfl,Abbott:2020mjq} consists of a BH within the upper mass gap where stellar evolution theories predict no black hole formation due to (pulsational) pair-instability supernovae \citep{2017ApJ...836..244W}. The total mass of this event is $\sim150M_\odot$, which is the most massive stellar-mass BBH system currently known.

\begin{figure*}
\begin{center}
\includegraphics[width=\linewidth]{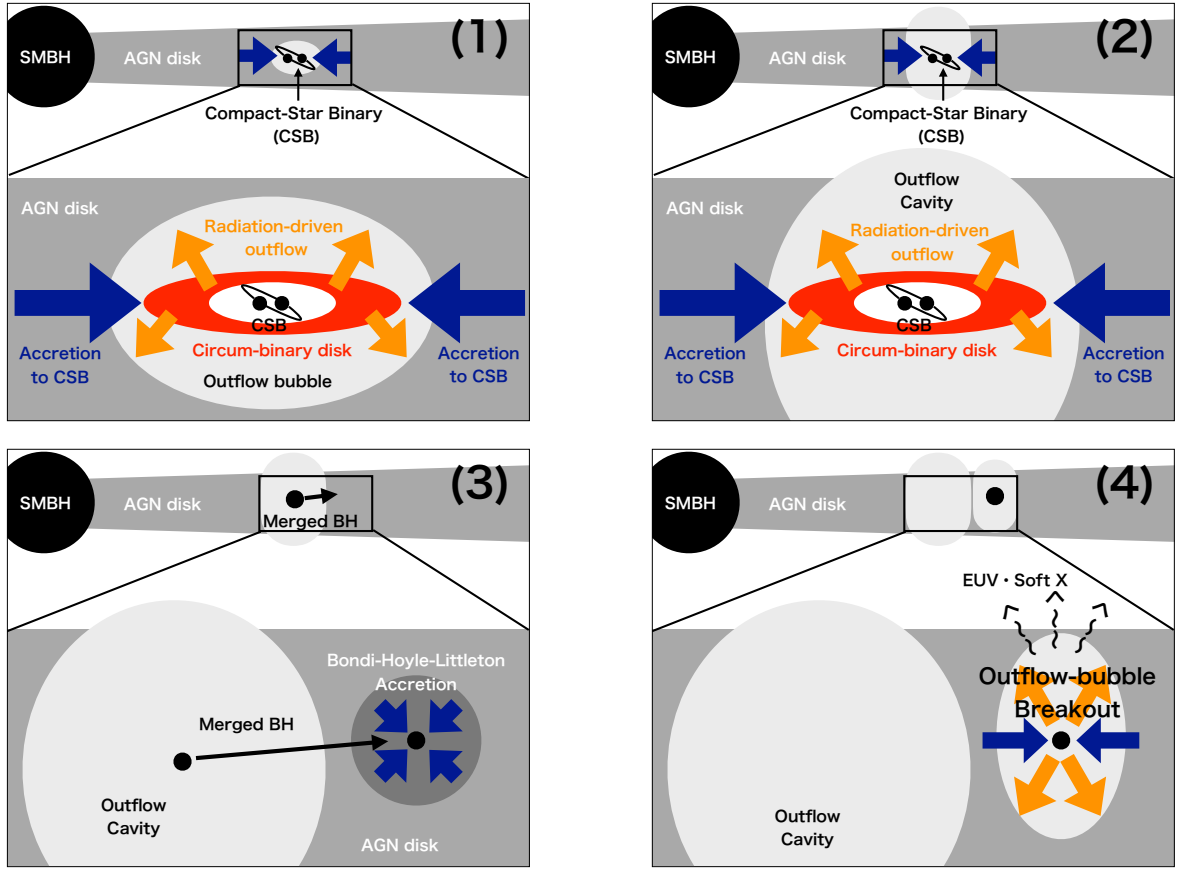}
\caption{{\bf Schematic picture of the evolution of outflows from a CSB embedded in an AGN disk.} (1) Gas in an AGN disk accretes onto a CSB. A circum-binary disk is formed due to the angular momentum transport via the shear motion. Due to the high accretion rate, the circum-binary disk produces radiation-driven outflows, leading to the formation of an outflow bubble. \added{We expect that outflows are mainly launched to the vertical direction, while the accretion proceeds in the midplane. Such a configuration enables the CSB to continuously accrete the AGN disk gas even in the outflow bubble.}
(2) The bubble expands and eventually punches out the AGN disk, making a cavity around the CSB. This typically happens before the binary merges. 
(3) The merger recoils the remnant BH that travels out of the cavity and into the dense AGN disk. As the BH reenters the AGN disk, it begins Bondi-Hoyle-Littleton accretion at a highly super-Eddington rate. 
(4) The radiation-driven outflows from the remnant BH penetrate the AGN disk, and produce the outflow-breakout emissions that outshine the AGN radiation in soft X-ray bands. 
\label{fig:schematic}}
\end{center}
\end{figure*}

These events are not expected by standard formation scenarios of merging BBHs, such as isolated binary evolution \citep{BHB16a,KIH14a}, and globular clusters \citep{RHC16a,FTM17a}. In particular, the formation of GW190521 is challenging, because both the primary and the secondary BHs are likely too massive (but for possible explanations, see \citealt{2020arXiv201007616T,2020arXiv200906585F,2020ApJ...903L..40L,2020ApJ...903L..21S} by Pop-III stars, \citealt{2020arXiv200913526B,2020arXiv201002242C} by uncertainty in the nuclear burning cross section,\added{ \citealt{2020arXiv201011730V} by a low-metallicity stellar evolution}).

Alternatively, nuclear star clusters in galaxies that host active galactic nuclei (AGNs) have been proposed as a possible BBH formation channel \citep{2012MNRAS.425..460M,BKH17a,SMH17a,2020ApJ...898...25T,2020arXiv201009765S}. In this scenario, interaction with the accretion disk aligns some of the BHs' orbits with the disk, after which the BHs migrate inwards within the disk. As black holes are compressed to an even smaller volume, they can undergo in multiple consecutive, so-called hierarchical, mergers, resulting in heavier black holes some of which can reside in the upper mass gap \citep{2019PhRvL.123r1101Y,2020ApJ...890L..20G,2020ApJ...898...25T,2020arXiv201200011T}. Similarly, hierarchical mergers involving neutron stars in AGN disks can result in merging objects in the lower mass gap \citep{2020ApJ...901L..34Y}. In addition, \cite{2020arXiv200905461G} reported that the gravitational waveform of GW190521 points to a highly eccentric merger, further supporting the event's dynamical/AGN origin \citep{2020arXiv201009765S,2020arXiv201010526T}.

\citet{2020PhRvL.124y1102G} recently reported an optical counterpart candidate to GW190521. The host galaxy of the counterpart is an AGN, and the claim is that the Bondi-Hoyle-Lyttleton accretion onto the merged BH powered the counterpart \citep{McKernan_2019}. Compact binary mergers involving neutron stars have also been proposed as possible multi-messenger sources within AGN disks. Recently, \citet{2020arXiv201108428Z} discussed the cases for binary neutron star and neutron star--black hole mergers and jet breakout emission from kilonova ejecta. \cite{2021ApJ...906L...7P} focused more generally on explosions in AGN disks and the ensuing breakout emission.

Outflow-driven transients and 
EM counterparts have been studied in the context of GW sources. If the accretion rate onto a BH is higher than the Eddington rate, radiation-driven outflows are produced~\citep{OMN05a,SNP13a,JSD14a}.
\cite{2016ApJ...822L...9M} proposed outflow-driven optical and radio transients powered by BBH mergers with mini-disks. \cite{2017ApJ...851...52K,2017ApJ...851...53K} investigated EM counterparts powered by sub-relativistic outflows at the secondary explosion in CSBs, including those induced by the Bondi-Hoyle-Lyttleton accretion onto the primary BH.
Disk-driven outflows are also relevant for the post-merger jet propagation, as discussed in the context of EM and neutrino counterparts of supermassive BH (SMBH) mergers~\citep{Yuan:2020oqg,Yuan:2021jjt}. 

In this paper, we consider radiation-driven outflows powered by the circum-binary disk formed around CSBs, which unavoidably affects the fate of post-merger outflows. We show a schematic picture of our scenario in Fig.~\ref{fig:schematic}.
Using the current understanding of accretion and outflow production processes, which have been mainly developed in the contexts of planet formation and black-hole accretion, respectively, we show that radiation-driven outflows produce outflow-bubbles inside AGN disks (see the panel 1 in Fig.~\ref{fig:schematic}). The outflows are so powerful that they can penetrate the AGN disk, forming a cavity around the CSB before the merger event in most of the suitable parameter range (panel 2 in Fig. \ref{fig:schematic})~\footnote{Density gaps can be formed by AGN disk-binary interactions, which also decreases the ambient density (see Section \ref{sec:mdot}). However, the gap density does not significantly decrease for most of the parameter space. A cavity has a much lower density than the gap.}. This cavity has such a low density that the merged BH cannot appreciably accrete from the surrounding medium {\it as long as it is in the cavity}. If the merged BH is kicked out of the cavity and into the intact AGN disk, then it can again accrete the surrounding gas at the Bondi-Hoyle-Littleton rate (panel (3) in Fig.~\ref{fig:schematic}). In this case, radiation-driven outflows are produced, and the outflow bubble breaks out the AGN disk again. Such an outflow-bubble breakout may emit detectable soft X-rays (panel (4) in Fig. \ref{fig:schematic}). 

This paper is organized as follows. We estimate mass accretion rates onto CSBs in AGN disks in Section~\ref{sec:mdot}. Then, conditions for outflow cavity formation are shown in Section~\ref{sec:cavity}. Our scenario for EM counterparts to BBH mergers are described in detail in Section~\ref{sec:EM}. We provide a summary, implications, and future prospects of our results in Section \ref{sec:summary}. We use the notation of $Q_X=Q/10^X$ in cgs unit except for masses of SMBHs and CSBs for which we use $M_\odot$.

\section{Accretion rates onto compact-stellar binaries}\label{sec:mdot}
\begin{figure}
\begin{center}
\includegraphics[width=\linewidth]{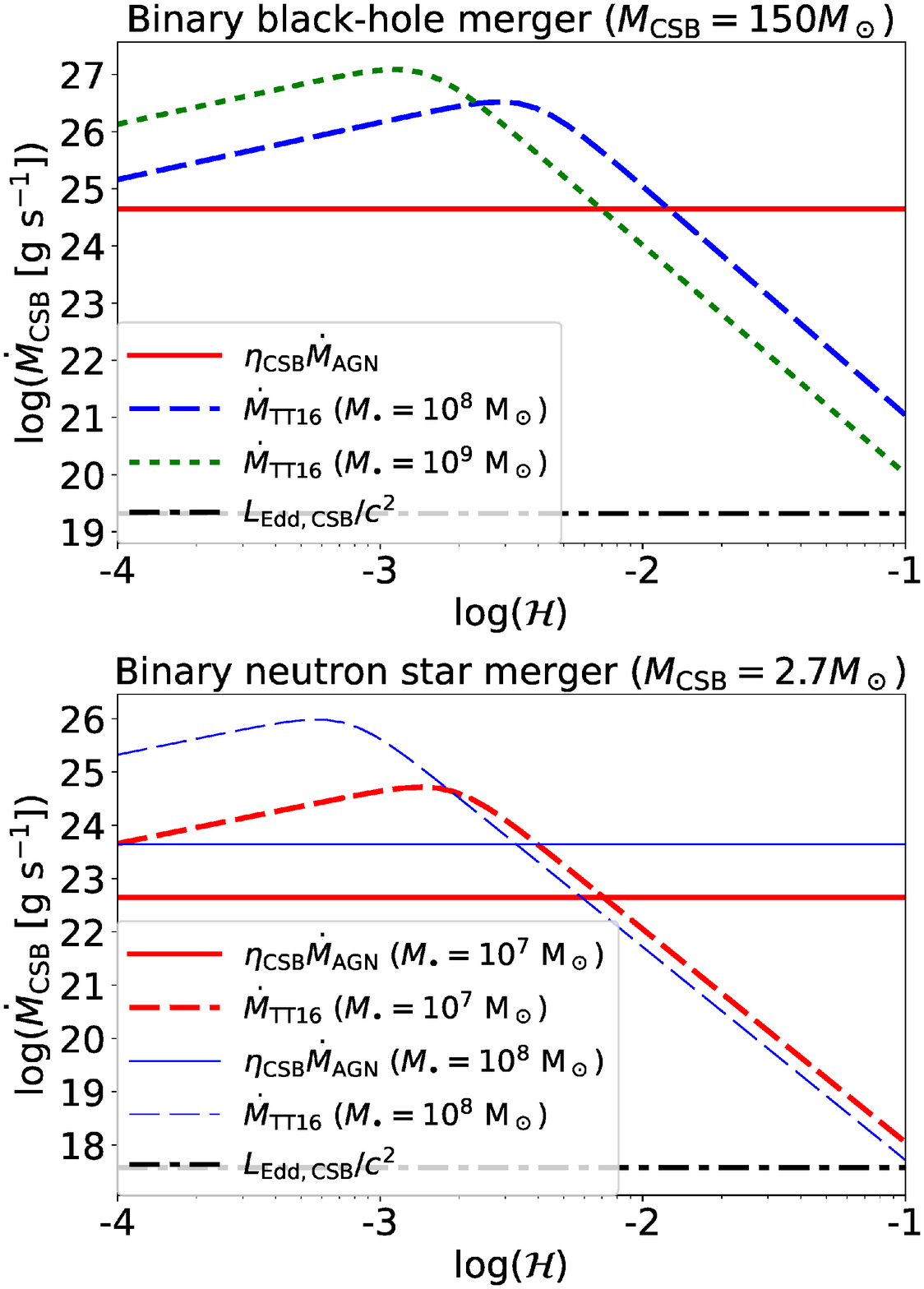}
\caption{{\bf Mass accretion rate onto a CSB in an AGN disk as a function of the aspect ratio, $\mathcal{H}$}. Upper panel: cases with $(M_{\rm CSB}/{M_\odot},~\dot{m}_{\rm AGN},~M_{\bullet}/{M_\odot})=(150,~0.2,~10^9)$ and $(150,~2.0,~10^8)$, which are motivated by the properties of GW190521 \citep{Abbott:2020tfl,Abbott:2020mjq} and its EM counterpart candidate \citep{2020PhRvL.124y1102G}. For both cases, the mass accretion rates to SMBHs are the same, $\dot{M}_{\rm AGN}c^2=2.5\times10^{46}\rm~erg~s^{-1}$. Lower panel: neutron-star mergers in typical AGNs, with $(M_{\rm CSB}/{M_\odot},~\dot{m}_{\rm AGN},~M_{\bullet}/{M_\odot})=(2.7,~1.0,~10^7)$ and $(2.7,~1.0,~10^8)$. Other parameters are $\alpha=0.1$, $\eta_{\rm CSB}=0.32$. $\dot{M}_{\rm TT16}$ is the mass accretion rates estimated by Equation (\ref{eq:MdotTT}) given in \citet{2016ApJ...823...48T}, while $\dot{M}_{\rm CSB}$ is limited by the mass supply rate from the outer AGN disk, $\eta_{\rm CSB}\dot{M}_{\rm AGN}$. The lower one of the two is realized. The mass accretion rates are highly super-Eddington for $\mathcal{H}\lesssim0.03$. }
\label{fig:mdot}
 \end{center}
\end{figure}

We consider an equal-mass CSB of total mass $M_{\rm CSB}$ and separation $a$ in an AGN disk surrounding a SMBH of mass $M_{\bullet}$ with an accretion rate $\dot{M}_{\rm AGN}=\dot{m}_{\rm AGN}L_{\rm Edd,AGN}/c^2\simeq1.4\times10^{25}~\dot{m}_{\rm AGN,0}M_{\bullet,8}\rm~g~s^{-1}$, where $L_{\rm Edd,AGN}\simeq1.3\times10^{46}~M_{\bullet,8}\rm~erg~s^{-1}$ is the Eddington luminosity, and $\dot{m}_{\rm AGN}$ is the normalized accretion rate. 
The CSB is located at $R=\mathcal{R}R_G=\mathcal{R}GM_{\bullet}/c^2\simeq1.5\times10^{16}~M_{\bullet,8}\mathcal{R}_3\rm~cm$ from the SMBH. We define the mass ratio $q\equiv M_{\rm CSB}/M_{\bullet}=10^{-6}~M_{\rm CSB,2}M_{\bullet,8}^{-1}$. 

We consider a viscous accretion disk using the $\alpha$ prescription, in which the radial velocity is estimated to be $V_R\approx \alpha \mathcal{H}^2 V_K\simeq9.5\times10^2~\alpha_{-1}\mathcal{H}_{-2.5}^2\mathcal{R}_3^{-1/2}\rm~cm~s^{-1}$, where $V_K=\sqrt{GM_\bullet/R}\simeq9.5\times10^8~\mathcal{R}_3^{-1/2}\rm~cm~s^{-1}$ is the Kepler velocity, $\alpha\sim0.1$ is the viscous parameter, $\mathcal{H}=H/R$ is the aspect ratio of the accretion disk,
\replaced{and $H$ is the disk scale height.}
{$H\approx (C_s/V_K)R$ is the disk scale height, and $C_s$ is the sound velocity in the disk.}
The scale height and aspect ratio should consistently be determined by the thermal balance and hydrostatic equilibrium.  In a standard viscous accretion disk, the aspect ratio does not strongly depends on any parameters \citep{ss73,KFM08a}, and we expect $10^{-3}\lesssim\mathcal{H}\lesssim3\times10^{-3}$ for the gas-pressure dominant regime. 
At the outer part of the AGN disk, the disk is gravitationally unstable, i.e., $Q=C_s\Omega/(\pi G\Sigma_{\rm AGN})\sim1$ \citep{Too64a},
\replaced{which induces star-formation activities. }{where $\Sigma_{\rm AGN}\approx\dot{M}_{\rm AGN}/(2\pi RV_R)\simeq1.6\times10^5~\dot{M}_{\rm AGN,25}R_{16}^{-1}V_{R,3}^{-1}\rm~g~cm^{-2}$ is the surface density of the AGN disk \citep{pri81}. The gravitational instability induces star-formation activities.}
Then, the feedback from massive stars heats up the gas, which likely maintains the disk marginally stable, $Q\sim1$. This can lead to a high value of $\mathcal{H}\sim0.1$ \citep{2005ApJ...630..167T,SMH17a}, although this mechanism can result in a relatively thin disk of $\mathcal{H}\lesssim0.01$ depending on parameters \citep{2020ApJ...898...25T}. Here, we provide $\mathcal{H}$ as a parameter, which allows us to investigate a wider parameter space.

The CSB accretes gas from the AGN disk, which is analogous to the gas accretion onto massive planets embedded in protoplanetary disks. 
\replaced{Based on the numerical simulations, the mass accretion rate is estimated to be \citep{2016ApJ...823...48T}}
{\cite{2016ApJ...823...48T} compiled 2-d and 3-d simulation results for the mass accretion process onto a planet embedded in a protoplanetary disk \citep{2002ApJ...580..506T,2003ApJ...599..548D,2010MNRAS.405.1227M}, and found that the mass accretion rate is well described by}
\begin{eqnarray}
\dot{M}_{\rm CSB}&\approx&\dot{M}_{\rm TT16}\approx0.3\mathcal{H}^{-2}q^{4/3}RV_K\Sigma_{\rm CSB}\label{eq:MdotTT}\\
&&\simeq3.0\times10^{26}~\mathcal{H}_{-2.5}^{-2}q_{-6}^{4/3}R_{16}V_{K,9}\Sigma_{\rm CSB,5}\rm~g~s^{-1}, \nonumber
\end{eqnarray}
where $\Sigma_{\rm CSB}$ is the surface density of the AGN disk at the position of the CSB.
\added{The parameter dependence of Equation (\ref{eq:MdotTT}) is consistent with the simple formula, $\dot{M}_{B,H}\approx \pi r_Br_{\rm Hill} C_s\rho_{\rm CSB}\propto \mathcal{H}^{-2}q^{4/3}RV_K\Sigma_{\rm CSB}$, where $r_B=2GM_{\rm CSB}/C_s^2\approx 2q\mathcal{H}^{-2}R$ is the Bondi radius, $r_{\rm Hill}=(q/3)^{1/3}R$ is the Hill radius, $\rho_{\rm CSB}\approx\Sigma_{\rm CSB}/(2H)$ is the density of the AGN disk, and we use $C_s\approx \mathcal{H}V_K$. Some previous studies utilized $\dot{M}_{\rm B,H}$ in the regime of $r_{\rm Hill}<H$ and $r_B<H$ \citep[e.g.][]{SMH17a,2020ApJ...898...25T}. Remarkably, $\dot{M}_{\rm CSB}\sim\dot{M}_{\rm B,H}$ is satisfied even for the regime of $r_B>H$ and $r_{\rm Hill}>H$ according to the simulations. Also, the relation is applicable for both $r_B > r_{\rm Hill}$ and $r_B<r_{\rm Hill}$. However, the simulation results slightly deviate from the values obtained by Equation (\ref{eq:MdotTT}) for low and high values of planet masses, based on Figure 1 in \citet{2016ApJ...823...48T}. Thus, it is unclear whether we can use the formula for all the parameter space. Future simulation studies with a wider parameter range might find the parameter space where another formula, such as the Bondi accretion rate, should be used.
}

\replaced{The surface density of the unperturbed AGN disk can be written as $\Sigma_{\rm AGN}\approx\dot{M}_{\rm AGN}/(2\pi RV_R)\simeq1.6\times10^5~\dot{M}_{\rm AGN,25}R_{16}^{-1}V_{R,3}^{-1}\rm~g~cm^{-2}$ \citep{pri81}. }
{The binary-AGN disk interaction can affect the surface density of the AGN disk. }
Because of the gravitational torque from the CSB, the density gap may ``open'' for a massive CSB, which results in $\Sigma_{\rm CSB}$ different from $\Sigma_{\rm AGN}$. Numerical simulations and analytic considerations of the planet-disk interaction process revealed that the surface density can be estimated to be \citep{2015MNRAS.448..994K}
\begin{equation}
    \Sigma_{\rm CSB}\approx{\rm min}(1,\chi_{\rm gap})\Sigma_{\rm AGN},
\end{equation}
where $\chi_{\rm gap}\approx32\mathcal{H}^{5}q^{-2}\alpha$. For our fiducial parameter set, we have $\chi_{\rm gap}\simeq1.0\mathcal{H}_{-2.5}^{5}q_{-6}^{-2}\alpha_{-1}$. 
\added{The gap opening corresponds to $\chi_{\rm gap}<1$, which occurs for a massive CSB or a geometrically thin AGN disk. Substituting the expressions for $R$, $V_K$, and $\Sigma_{\rm CSB}$, we obtain the parameter dependence of the mass accretion rate as $\dot{m}_{\rm TT16}=\dot{M}_{\rm TT16}/L_{\rm Edd,CSB}\propto\dot{m}_{\rm AGN}\alpha^{-1}\mathcal{H}^{-4}q^{1/3}$ for $\chi_{\rm gap}>1$ and  $\dot{m}_{\rm TT16}\propto\dot{m}_{\rm AGN}\mathcal{H}q^{-5/3}$ for $\chi_{\rm gap}<1$, where $L_{\rm Edd,CSB}$ is the Eddington luminosity for the CSB. The normalization of $\dot{M}_{\rm CSB}$ is given in Equation (1) with $\chi_{\rm gap}\simeq1$. }

If $\dot{M}_{\rm TT16}$ is higher than $\dot{M}_{\rm AGN}$, the mass accretion onto the CSB is simply limited by the mass supply from the outer AGN disk. Then, the accretion rate is written as
\begin{eqnarray}
\dot{M}_{\rm CSB}&\approx&\eta_{\rm CSB}\dot{M}_{\rm AGN}\\
 &\simeq&1.4\times10^{24}~\dot{m}_{\rm AGN,0}M_{\bullet,8}\eta_{\rm CSB,-1}\rm~g~s^{-1}\nonumber
\end{eqnarray}
where $\eta_{\rm CSB}<1$ is a parameter that describes a fraction of AGN disk mass transferred to the CSB. The value of $\eta_{\rm CSB}$ is uncertain, although 2D simulations may suggest $\eta_{\rm CSB}\sim0.5$ \citep{2021arXiv210109406L}.
\replaced{Therefore,}{ Combining the two regimes,}
the mass accretion rate onto the CSB is represented as $\dot{M}_{\rm CSB}=\min(\dot{M}_{\rm TT16},~\eta_{\rm CSB}\dot{M}_{\rm AGN})$.
Interestingly, the mass accretion rate is independent of $\mathcal{R}$ in all the branches.
\added{One may think that the SMBH would be starved unless $\eta_{\rm CSB}\ll1$, since there are many CSBs embedded in an AGN disk. However, for the cases with $\dot{M}_{\rm CSB}=\eta_{\rm CSB}\dot{M}_{\rm AGN}$, the outflow velocity should be lower than the escape velocity of the SMBH, because the outflow production radius (see Section \ref{sec:cavity}) is large for $\eta_{\rm CSB}\gtrsim0.1$. Then, the outflows will fall back to the AGN disk, and thus, the SMBH is not starved in our scenario.}

We plot the mass accretion rate as a function of $\mathcal{H}$ for parameter sets for GW190521 and binary neutron-star mergers in typical AGN (see captions for other parameter sets).
We can see that the mass accretion rate is limited by $\dot{M}_{\rm AGN}$ for $\mathcal{H}\lesssim0.01$, where we see that $\dot{M}_{\rm CSB}c^2\gg L_{\rm Edd,CSB}$. Such a high mass accretion rate leads to production of powerful radiation-driven outflows \citep{OMN05a,SNM14a,JSD14a,TOK16a}. For a further lower value of $\mathcal{H}\lesssim10^{-3}$, the density gap opens up in the AGN disk due to the binary-disk interactions. This leads to a low value of $\dot{M}_{\rm TT16}$, but the mass accretion rate is still determined by $\eta_{\rm CSB}\dot{M}_{\rm AGN}$ and highly super-Eddington in the reasonable range of $\mathcal{H}$. For an opposite limit of $\mathcal{H}\gtrsim0.01$, $\dot{M}_{\rm CSB}=\dot{M}_{\rm TT16}$ is satisfied. The mass accretion rate onto the CSB is lower for a higher $\mathcal{H}$, and close to the Eddington value at $\mathcal{H}\sim0.1$ for all the cases.

\section{Cavity formation by radiation-driven outflows}\label{sec:cavity}
\begin{figure*}
\begin{center}
\includegraphics[width=\linewidth]{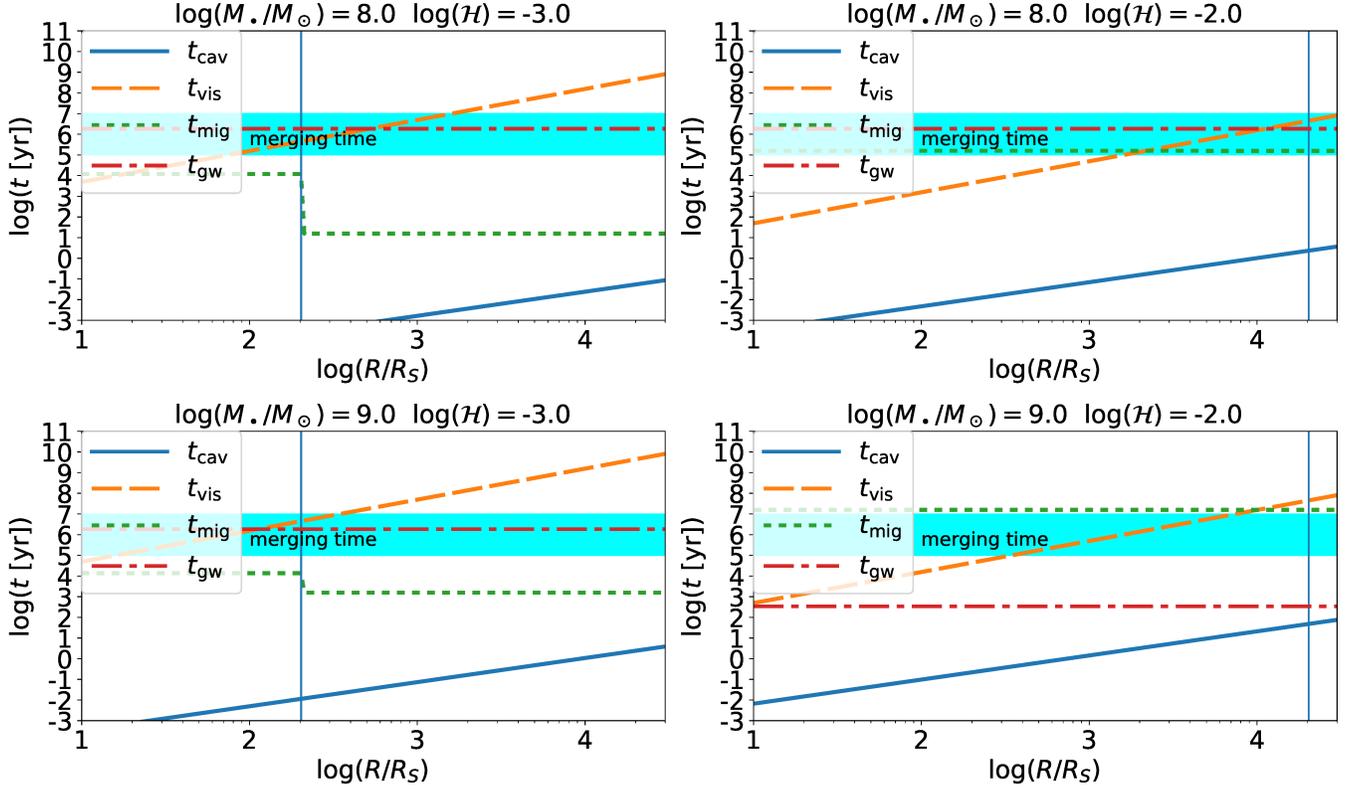}
\caption{{\bf Comparisons of timescales as a function of $\mathcal{R}=R/R_S$.} The top-left, top-right, bottom-left, and bottom-right panels are for ($M_{\bullet},~\mathcal{H})=(10^8,~10^{-3})$, $(10^8,~10^{-2})$, $(10^9,~10^{-3})$, $(10^9,~10^{-2})$, respectively. The shaded region represents the timescale for the CSB merger in AGN disks \citep{2020ApJ...898...25T}. The vertical lines represent the critical radius above which the AGN disk is gravitationally unstable. The values of the other parameters are $M_{\rm CSB}=150\rm~M_\odot$, $\eta_w=0.32$ $\eta_{\rm CSB}=0.32$, $\alpha=0.1$, $v_w=10^9\rm~cm~s^{-1}$, and $A_{\rm in}=11$. We can see that the cavity formation timescale is the shortest in all the panels.  \label{fig:timescale}}
\end{center}
\end{figure*}

\begin{figure*}
\begin{minipage}{0.33\hsize}
\begin{center}
\includegraphics[width=\linewidth]{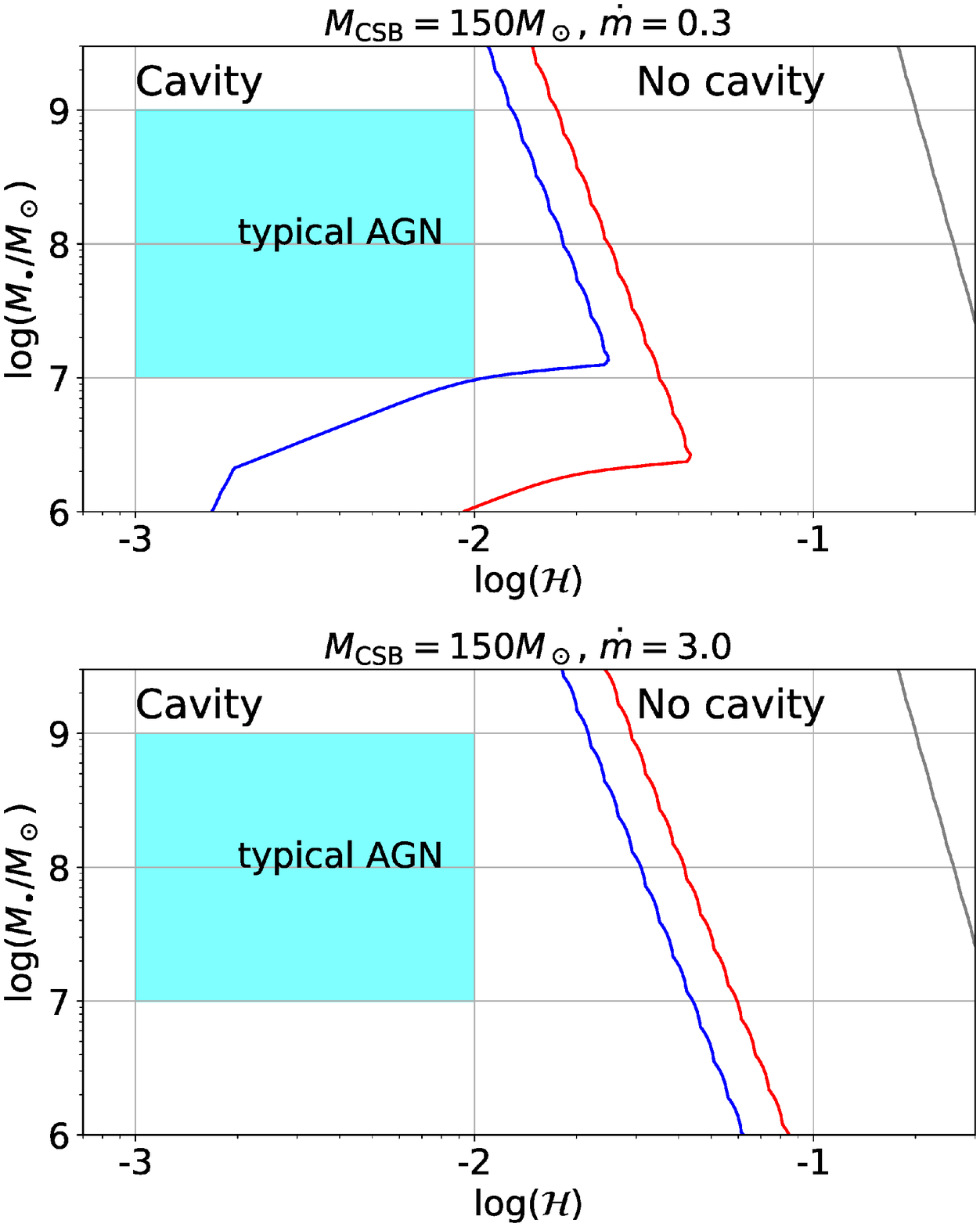}
\end{center}
\end{minipage}
\begin{minipage}{0.33\hsize}
\begin{center}
\includegraphics[width=\linewidth]{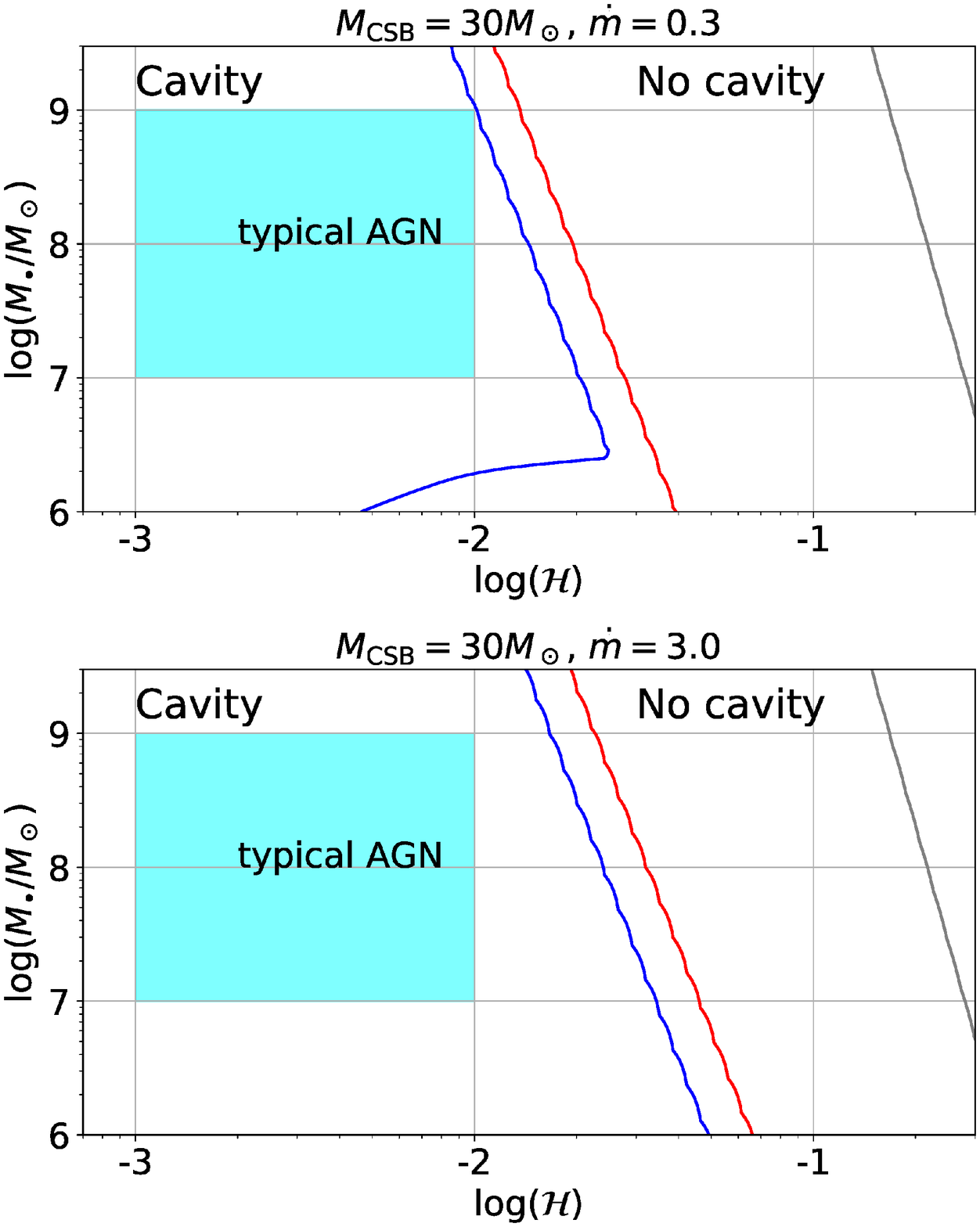}
\end{center}
\end{minipage}
\begin{minipage}{0.33\hsize}
\begin{center}
\includegraphics[width=\linewidth]{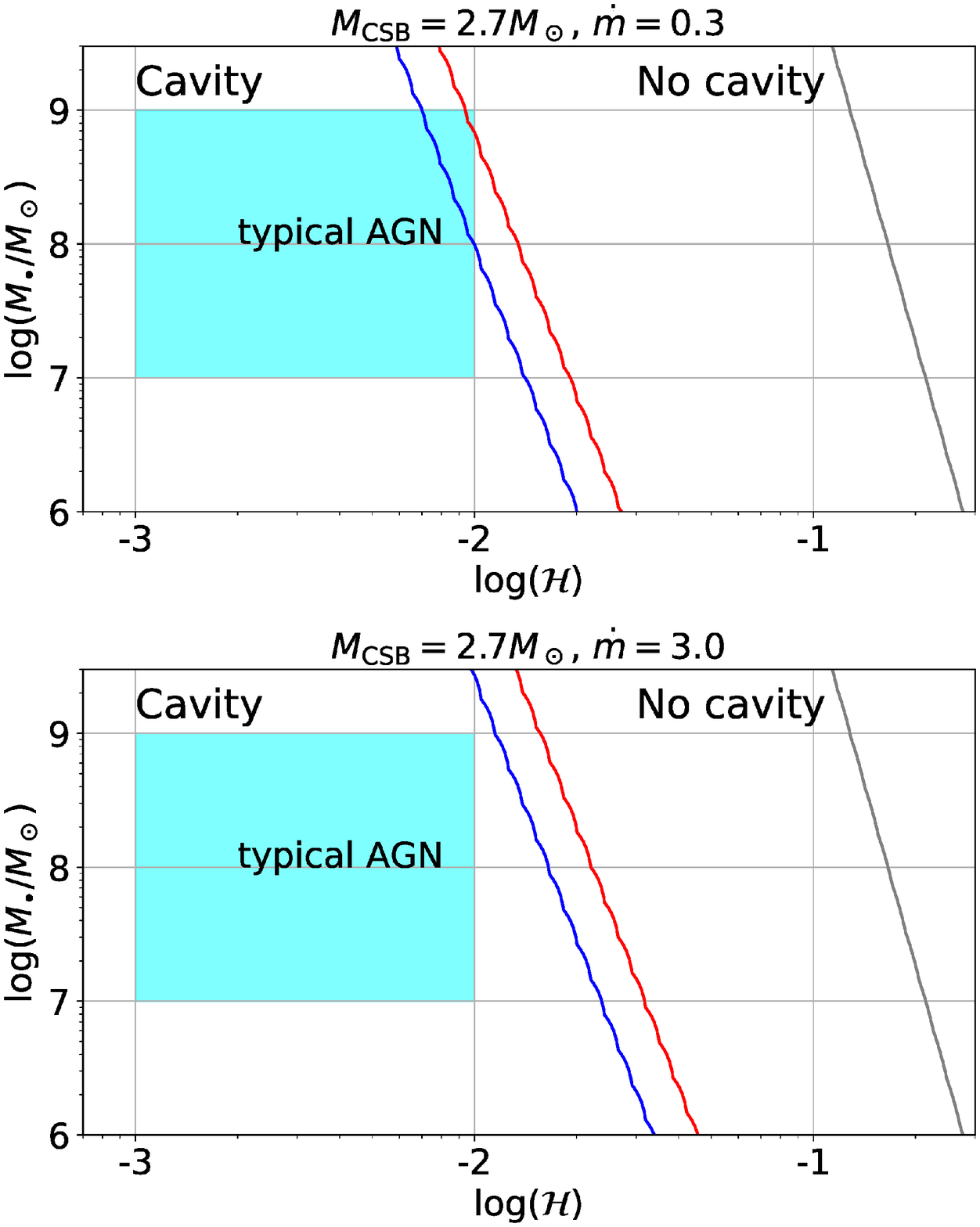}
\end{center}
\end{minipage}  
\caption{{\bf Cavity formation conditions in $\mathcal{H}-M_{\bullet}$ plane.} The left, middle, and right panels are for the cases with a GW190521-like event, a typcal BBH merger, and a typical BNS merger, respectively. The top and bottom panels are for $\dot{m}_{\bullet}=0.3$ and 3.0, respectively. The blue and red lines are the cavity formation conditions by circum-binary outflows at $\mathcal{R}=10^4$ and $10^2$, respectively. The gray lines show the cavity formation condition by mini-disk-driven outflows. The cyan shaded regions are the parameter space expected for typical AGNs \added{(e.g., \citealt{lhw11,Ueda:2014tma} for the SMBH mass and \citealt{ss73,KFM08a} for the disk scale height in the standard disk)}, which lies in the cavity formation regime. Other parameters are the same as those in Fig.~\ref{fig:timescale}.   \label{fig:cavity_param}}
\end{figure*}

Because of the shear motion of the AGN disk, the accreting gas has an angular momentum which is aligned to that of the AGN disk~\citep{1999ApJ...526.1001L,2002ApJ...580..506T}. The accreting gas is circularized at \citep{2012ApJ...747...47T}
\begin{equation}
r_{\rm circ}\approx0.1r_{\rm Hill}\simeq 1.0\times10^{13}q_{-6}^{1/3}\mathcal{R}_3M_{\bullet,8}\rm~cm,   
\end{equation}
where \deleted{$r_{\rm Hill}= (q/3)^{1/3}R$ is the Hill radius and} the factor 0.1 is calibrated by hydrodynamic simulations. The circularized gas forms a circum-binary disk. For a merging CSB, $r_{\rm circ}$ is much larger than the binary separation, $a$. Then, the gravitational force of the CSB exerted on the circum-binary disk is approximated by a point source, and the evolution of the circum-binary disk is described by the theory of accretion flows onto a single BH as long as $r\gg a$.  

\subsection{Outflows from circum-binary disks}
\replaced{Since the accretion rate is highly super-Eddington, the circum-binary outflows are produced if the inner radius of the circum-binary disk is less than the outflow-production radius, $r_w\approx\dot{m}_{\rm CSB}r_G$, where  $\dot{m}_{\rm CSB}=\dot{M}_{\rm CSB}c^2/L_{\rm Edd,CSB}$ is the normalized accretion rate \citep{ss73,KFM08a}.}
{An accretion flow with a highly super-Eddington rate produces outflows using the radiation pressure. The outflows are expected to be produced at the point where the accretion luminosity becomes higher than the Eddington luminosity, i.e., $GM_{\rm CSB}\dot{M}_{\rm CSB}/r_w\approx L_{\rm Edd,CSB}$. This leads to the expression of the outflow production radius of $r_w\approx\dot{m}_{\rm CSB}r_G$, where $r_G=GM_{\rm CSB}/c^2$. At the vicinity of the CSB, the circum-binary disk is torn apart via interactions with the CSB}. 
The inner edge of the circum-binary disk is determined by the balance between the precession torque from the binary and the viscous torque in the disk, which leads to \citep{2013MNRAS.434.1946N}
\begin{equation}
 r_{\rm in}\approx A_{\rm in}a\simeq 16 \mu^{1/2}(\sin2\theta)^{1/2}(h/r)_{-2}^{-1/2}\alpha_{-1}^{-1/2}a,
\end{equation}
where $\mu=M_{\rm sec}/(M_{\rm pri}+M_{\rm sec})$ is the binary mass ratio, $\theta$ is the angle between the binary orbital plane and the circum-binary disk, and $(h/r)$ is the aspect ratio of the circum-binary disk. We expect outflows when $r_w>r_{\rm in}$, i.e., $a<a_w=\dot{m}_{\rm CSB}r_G/A_{\rm in}$. 
\added{On the other hand, we do not expect the outflows from the circum-binary disk for $r_w < r_{\rm in}$. In this case, the CSB accretes the gas through the mini-disks surrounding each BH. We will discuss this situation in Subsection \ref{sec:minidisk}. }

The duration of outflow production from circum-binary disks is limited by the timescale of the CSB merger. Since the separation is close enough when the outflows are produced, GW radiation is the dominant process of binary separation in most cases. Then, we estimate the outflow duration to be \citep[e.g.,][]{ST83a} 
\begin{eqnarray}
t_{\rm gw}&=&\frac{5}{128}\frac{c^5a_w^4}{G^3M_{\rm CSB}^3}\approx\frac{5}{128}\frac{\dot{m}_{\rm CSB}^4}{A_{\rm in}^4}\frac{r_G}{c}\\
&\simeq&1.9\times10^{11}~\dot{m}_{\rm CSB,5}^4A_{\rm in,1}^{-4}M_{\rm CSB,2}{\rm~s},\nonumber
\end{eqnarray}
\replaced{When $t_{\rm gw}$ is longer than Myr, other processes, such as binary-single interactions, can limit the duration of the outflows from the circum-binary disk.}
{where we use $r_G=GM_{\rm CSB}/c^2$ and $a_w=\dot{m}_{\rm CSB}r_G/A_{\rm in}$ in the second equation.
The binary-single interactions happening in AGN disks can determine the merger timescale if $t_{\rm gw}\gtrsim0.1-1$ Myr \citep{2020ApJ...898...25T}. The binary-single interactions occur using the difference of the migration velocity between the CSB and the third body, and several binary-single interactions can lead to a merger event. With a typical parameters, the merger timescale by the binary-single interactions is $10^5 - 10^7$ yr \citep{2020ApJ...898...25T}. AGN disk-binary interactions may also affect the merger timescale, which also leads to a typically merger timescale of the order of Myr \citep[e.g.][]{SMH17a}.}

The radiation-driven outflows create a wind bubble as in the surrounding of massive stars. For a uniform density, the bubble expands with time as $r_{\rm bub}\approx0.88(L_wt^3/\rho_{\rm CSB})^{1/5}$ \citep{1977ApJ...218..377W,1992ApJ...388...93K}, where $L_w=\eta_w\dot{M}_{\rm CSB}v_w^2\simeq3.2\times10^{41}~\dot{M}_{\rm CSB,24}\eta_{w,-0.5}v_{w,9}^2\rm~erg~s^{-1}$ is the kinetic luminosity of the outflows, $\eta_w$ is the outflow production efficiency, $v_w$ is the outflow velocity, and $\rho_{\rm CSB}=\Sigma_{\rm CSB}/(2H)\simeq1.6\times10^{-9}\Sigma_{\rm CSB,5}H_{13.5}^{-1}~\rm~g~cm^{-3}$ is the mass density in the AGN disk at the position of the CSB. Radiation hydrodynamic simulations suggest that $\eta_w>0.9$ for highly super-Eddington accretion of $\dot{m}_{\rm CSB}\gtrsim10^4$ \citep{2015ApJ...806...93J,2018PASJ...70..108K}. Nevertheless, we conservatively use $\eta_w=0.32$ as a fiducial value, which is suitable for $\dot{m}_{\rm CSB}\sim10^2$ \citep{JSD14a}. The bubble continuously expands, and the outflow bubble penetrates the AGN disk and make a cavity in the disk within a timescale of 
\begin{eqnarray}
t_{\rm cav}&\approx&\left(\frac{\rho_{\rm CSB}H^5}{0.53L_w}\right)^{1/3}\label{eq:tcav}\\
& &\simeq5.7\times10^5~\rho_{\rm CSB,-9}^{1/3}H_{13.5}^{5/3}L_{w,41.5}^{-1/3}~\rm s. \nonumber
\end{eqnarray}
\added{Here, we assume that the outflow luminosity is independent of the bubble size. Feedback by the outflow bubbles may affect the mass accretion rate and the outflow luminosity. The feedback is actively discussed in the context of the growth of SMBHs \citep{2007ApJ...661..693P,2009ApJ...698..766M}, and the simulations with anisotropic feedback result in the accretion rate comparable to the Bondi rate \citep[e.g.][]{2017MNRAS.469...62S,2018MNRAS.476..673T}, supporting our assumption. Future studies with parameter sets for stellar-mass BHs will be able to quantitatively understand $\dot{M}_{\rm CSB}$ with the feedback.}

If $t_{\rm cav}<t_{\rm gw}$, the outflow bubble penetrates the AGN disk, and a cavity is inevitably formed before the merger. 
Fig.~\ref{fig:timescale} shows $t_{\rm cav}$ and $t_{\rm gw}$ as a function of $\mathcal{R}$ for the cases with a GW190521-like event. Based on N-body simulations that include relevant processes, most of the BBH mergers occur for $\lesssim0.01$ pc \citep{2020ApJ...898...25T}. Thus, we plot the timescales for $\mathcal{R}<10^4$. We can see that $t_{\rm cav}$ is shorter in the range of our calculations, indicating that outflow cavities are created. 

We expand our investigation range for the cavity formation for various values of $M_{\rm CSB}$, $\dot{m}_{\rm AGN}$, $M_{\bullet}$, and $\mathcal{H}$. Fig.~\ref{fig:cavity_param} depicts the parameter space where cavity is formed in the $\mathcal{H}-M_{\bullet}$ plane. The cavity formation can be avoided only for high $\mathcal{H}$ cases. The mass accretion rate onto the CSB is strongly suppressed as $\dot{m}_{\rm CSB}\propto\mathcal{H}^{-4}$, which leads to small values of $t_{\rm gw}$ and large values of $t_{\rm cav}$. 
Setting $t_{\rm cav}=t_{\rm gw}$, a necessary condition for the cavity formation is given by $ \mathcal{H}\gtrsim\mathcal{H}_{\rm crit}$, where
\begin{equation}
\mathcal{H}_{\rm crit}\simeq0.04~q_{-6}^{25/162}\dot{m}_0^{2/9}\mathcal{R}_4^{-2/27}\alpha_{-1}^{-13/72}\eta_{w,-0.5}^{1/54}v_{w,9}^{1/54}A_{\rm in,1}^{-2/9}.\label{eq:Hcrit}
\end{equation}
Here, we use $\dot{M}_{\rm CSB}=\dot{M}_{\rm TT16}$ and the disk structure without a gap, i.e., $\chi_{\rm gap}>1$. We see that the parameter dependence of the critical aspect ratio is very weak, and thus, the cavity should be formed in the standard disk of $\mathcal{H}\lesssim0.03$. We stress that the outflow cavity is formed for a wide parameter range of AGN accretion disks.

If $M_{\rm CSB}$ is sufficiently high, or if $M_{\bullet}$ and/or $\dot{m}_{\rm AGN}$ are sufficiently low, the cavity formation may be avoided. The mass accretion rate for the CSB is limited by $\eta_{\rm CSB}\dot{M}_{\rm AGN}$ in this case, which makes $t_{\rm cav}$ longer and $t_{\rm gw}$ shorter. We see this in the left top panel in Figure \ref{fig:cavity_param}. For a low value of $\mathcal{H}$, the density gap is also formed, which changes the parameter dependence of the relevant timescales. This feature is also seen in the top left panel.

\subsection{Outflows from mini-disks}\label{sec:minidisk}
Next, we discuss the effect of mini-disks surrounding each compact object. The circum-binary disk is destroyed at $r\approx r_{\rm in}$, and the accreting gas forms two mini-disks around the primary and secondary, respectively. A circum-binary disk with a high accretion rate should have a large aspect ratio, $(h/r)\gtrsim0.1$. This makes a turbulent viscosity stronger than the torque exerted by the binary orbital motion. Then, most of the accretion gas can enter into the binary orbit and forms the mini disks. This picture is supported by the recent simulations, where the mass accretion rate in the mini disks are comparable to that in the circum-binary disk \citep{2013MNRAS.436.2997D,FDM14a,2019ApJ...875...66M}. Thus, the outflow rate from the mini-disks is likely not to be much different from that from the circum-binary disks, and we estimate the cavity production timescale by Equation (\ref{eq:tcav}).

The cavity production is interrupted by viscous diffusion of the AGN disk material, migration by the disk-CSB interaction, or the merger of the CSB. The viscous timescale of the AGN disk is estimated to be \citep{pri81}
\begin{equation}
t_{\rm vis}\approx\frac{R}{\alpha\mathcal{H}^2V_K}\simeq1.5\times10^{13}\mathcal{R}_3^{3/2}M_{\bullet,8}\alpha_{-1}^{-1}\mathcal{H}_{-2.5}^{-2}\rm~s.
\end{equation}
The migration timescale depends on the gravitational stability of the AGN disk and the existence of the density gap. For a gravitationally stable AGN disk, recent numerical simulations revealed that the migration timescales without a gap ($\chi_{\rm gap}>1$) and with a gap ($\chi_{\rm gap}<1$) are given by a simple formula \citep{2018ApJ...861..140K}:
\begin{eqnarray}
t_{\rm mig}&\approx&\frac{\mathcal{H}^2M_{\bullet}}{6qRV_K\Sigma_{\rm CSB}}\\
&\simeq&3.3\times10^{11}\mathcal{H}_{-2.5}^2M_{\bullet,8}q_{-6}^{-1}R_{16}^{-1}V_{K,9}^{-1}\Sigma_{\rm CSB,5}^{-1}\rm~s.\nonumber
\end{eqnarray}
Note that $\Sigma_{\rm CSB}$ depends on $\chi_{\rm gap}$ and $t_{\rm mig}$ is longer with a density gap.
If the AGN disk is gravitationally unstable, the migration timescale is estimated to be $t_{\rm mig}\approx\mathcal{H}^2M_{\bullet}/(6qRV_K\Sigma_{\rm AGN})$ \citep{2011MNRAS.416.1971B}, which is the same as the migration timescale of the gravitationally stable disk without a gap. The CSB merger timescale is determined by the binary-single interactions, which ranges from $t_{\rm mer}\sim10^5-10^7$ yr \citep{2020ApJ...898...25T}. Figure \ref{fig:timescale} also shows these timescales as a function of $\mathcal{R}$.  
For the range of our interest, $t_{\rm cav}$ is the shortest, and thus, the cavity is inevitably created before the merger.

With our formulation, $t_{\rm mig}$ does not depend on $\mathcal{R}$, while $t_{\rm cav}\propto\mathcal{R}^{7/6}$ without a density gap. We do not expect cavity formation for the radii where $t_{\rm mig}<t_{\rm cav}$ is satisfied. The CSB migrates inward with a timescale of $t_{\rm mig}$, and creates a cavity once it reaches a radius where $t_{\rm mig}>t_{\rm cav}$ is satisfied. Thus, the cavity formation condition should be evaluated by comparison of $t_{\rm cav}$ to $t_{\rm vis}$ and $t_{\rm mer}$. The gray lines in Fig.~\ref{fig:cavity_param} indicate the boundary above which cavity formation by mini-disk outflows can be avoided. Only the AGN disk with a very high aspect ratio, namely $\mathcal{H}>0.1$, can avoid the cavity formation. Since such a value is unexpected in a typical AGN disk, we conclude that the cavity formation is inevitable for the quasi-aligned binaries. This conclusion should be unchanged even for the mildly misaligned case, because the relevant timescales are identical as long as the mini-disk-driven outflows have a component perpendicular to the AGN disk.

In summary, cavity formation is avoided only if the orbital plane of the CSB is quasi-perpendicular to the AGN disk. Also the AGN disk should have a relatively high aspect ratio given by Equation (\ref{eq:Hcrit}). 

\section{Electromagnetic Counterparts from Mergers inside Outflow Cavities} \label{sec:EM}
As shown in the previous section, the cavity formation is highly likely for CSBs in AGN disks. 
The density of the outflow cavity is much lower than the AGN disk density, so that EM and neutrino counterparts of GW events inside the cavity are too dim to be observed. 
Possible post-merger jets (assuming that they are GRB-like) are unlikely to be choked by the AGN disk (contrary to the conclusion by \cite{2020arXiv201108428Z}), unless the jet direction is aligned with the AGN disk. Choked jets have been proposed as the sources of high-energy neutrinos~\citep[e.g.,][]{Murase:2013ffa,Senno:2015tsn,Tamborra:2015fzv,Kimura:2018vvz}, but we expect that such a system is much rarer than compact-star merger events inside the AGN disk.

In reality, the remnants of compact binary mergers with a significant asymmetry, in terms of mass or spin, will receive a recoil velocity upon merger due to GW radiation~\citep[see][for a review]{2010RvMP...82.3069C}. Such a recoil motion changes the dynamics of the surrounding gas, which may trigger EM transients~\citep[see][]{2008ApJ...676L...5L,MK17a,McKernan_2019,2020PhRvL.124y1102G}. 

In the following, we discuss scenarios of EM counterparts in detail (see Fig.~\ref{fig:schematic}). 
If the merged BH is kicked into the vertical direction, it moves inside the outflow cavity. Then, the mass accretion rate onto the kicked BH is very low, which results in essentially no optical or X-ray counterparts that outshine the AGN emission.
We hereafter show that detectable EM counterparts require some special conditions, implying that the rate density of GW events with EM counterparts would be much lower than that of all the merger events inside the AGN disk. 
We should keep in mind that the EM transients produced by the kicked BH should be as luminous as the host AGN in order to be identified as the EM counterparts. This provides a strong constraint on the detectability.

\subsection{Mass accretion onto the BHs kicked into the AGN disk}
If the merged BH is kicked ``along'' the AGN disk plane with a sufficiently high kick velocity, $v_{\rm kick}$, the merged BH can escape from the cavity and be kicked into the AGN disk. Usually, the kick velocity ($\sim10^2-10^3\rm~km~s^{-1}$; \citealt{2007PhRvL..98i1101G,2007CQGra..24S..33H,2007PhRvL..98w1102C,2007ApJ...659L...5C}) is less than the escape velocity ($\sim1\times10^4\mathcal{R}_3^{-1/2}\rm~km~s^{-1}$), and the kicked BH experiences the epicyclic motion.
The maximum radial displacement to the radial direction is given by $\delta R\approx (v_{\rm kick}/V_K)^2 R$. The size of the cavity, $r_{\rm cav}$, is expected to be comparable to the Hill radius, $r_{\rm Hill}\sim6.9\times10^{13}~R_{16}q_{-6}^{1/3}$ cm. If the cavity expands larger than the Hill radius or Bondi radius, $r_B=2GM_{\rm CSB}/C_s^2\sim2\times10^{15}~\mathcal{H}_{-2.5}^{-2}q_{-6}R_{16}$ cm, the accretion and outflows should stop. This may regulate the size of the cavity to be comparable to the smaller of the Hill radius and Bondi radius.  In the range of our interest, $r_B$ is always larger than $r_{\rm Hill}$, so we expect that the cavity radius is regulated to the order of the Hill radius, i.e., $r_{\rm cav}\sim r_{\rm Hill}$, which is also not far from the scale height $H\sim3.2\times10^{13}R_{16}~\mathcal{H}_{-2.5}$ cm for our typical parameter set. 
We write the condition that the merged BH can get into the AGN disk again by crossing the cavity as $r_{\rm cav}\lesssim\delta R$, or
\begin{eqnarray}
v_{\rm kick}\gtrsim v_{\rm kick,cr}&=&\left(\frac{r_{\rm cav}}{R}\right)^{1/2}V_K\\
&\simeq&4.4\times10^2~r_{\rm cav,13.5}^{1/2}M_{\bullet,8}^{-1/2}\mathcal{R}_3^{-1}\rm~km~s^{-1}.\nonumber
\end{eqnarray}
The critical velocity depends on $r_{\rm cav}$, which may be more close to $\sim1000~{\rm km}~{\rm s}^{-1}$ if $r_{\rm cav}\sim r_{\rm Hill}$. The above condition can be satisfied for a binary with a high spin. The kick velocity can be as high as $v_{\rm kick}\sim300\rm~km~s^{-1}$ without a spin~\citep{2007PhRvL..98i1101G,2007CQGra..24S..33H} and $v_{\rm kick}\sim4000\rm~km~s^{-1}$ with a high spin \citep{2007PhRvL..98w1102C,2007ApJ...659L...5C}. 
Indeed, BHs in GW190521 have a high spin of $a\sim0.7$ before the merger \citep{Abbott:2020tfl,Abbott:2020mjq}, which could be consistent with the value expected for a remnant BH after the merger \citep{2008PhRvD..78d4002R}. 

Once the merged BH enters into the AGN disk, the merged BH accretes the AGN disk gas. Owing to a high kick velocity, the accretion radius, $r_{\rm BHL}=2GM_{\rm CSB}/(C_s^2 + v_{\rm kick}^2)\approx2GM_{\rm CSB}/v_{\rm kick}^2\simeq2.7\times10^{13}~M_{\rm CSB,2}v_{\rm kick,7.5}^{-2}$ cm, can be smaller than the Hill radius and the scale height. Then, we can use the well-known formula for the estimate of the accretion rate \citep{HL39a,Bon52a,Edg04a}: 
\begin{eqnarray}
\dot{M}_{\rm BHL}&=&\frac{4\pi G^2M_{\rm CSB}^2\rho_{\rm CSB}}{v_{\rm kick}^3}\\ 
&\simeq&7.0\times10^{25}M_{\rm CSB,2}\rho_{\rm CSB,-9}v_{\rm kick,7.5}^{-3}\rm~g~s^{-1}.\nonumber
\end{eqnarray}
\added{The velocity shear in the AGN disk may reduce the mass accretion rate. We estimate the shear velocity to be $V_{\rm she}\approx(r_{\rm gap}/R)V_K\sim30 \mathcal{R}_3^{-1/2}(r_{\rm gap}/R)_{-2.5}\rm~km~s^{-1}$, which is much lower than $v_{\rm kick}$. Thus, the shear does not affect the accretion rate in our situation.}
The epicyclic motion determines the duration of the mass accretion process, which is 
\begin{equation}
t_K\approx\Omega_K^{-1}\simeq1.8\times10^2 ~M_{\bullet,8}\mathcal{R}_3^{3/2}\rm~day.
\end{equation}
The outflow luminosity is estimated to be $L_w\approx\eta_w \dot{M}_{\rm BHL}v_w^2\simeq3.2\times10^{45}\dot{M}_{\rm BHL,26}\eta_{w,-0.5}v_{w,10}^2\rm~erg~s^{-1}$.

We cautiously note that the mass accretion rate in this phase can exceed $\dot{M}_{\rm AGN}$, because the duration is much shorter than the AGN lifetime. Total mass that accretes onto the merged BH is much lower than the AGN disk mass there. Also, the density gap produced by the AGN disk-binary interaction still exists after the merger event. The gap will be filled in the viscous timescale of the gap width, $\sim R_{\rm gap}^2/(\alpha\mathcal{H}^2V_KR)\sim 3.0\times10^3 ~q_{-6} \alpha_{-1}^{-3/2}\mathcal{H}_{-2.5}^{-7/2}  t_K$, where $R_{\rm gap}/R\approx 0.41~q^{1/2}\mathcal{H}^{-3/4}\alpha^{-1/4}\simeq0.055~q_{-6}^{1/2}\mathcal{H}_{-2.5}^{-3/4}\alpha_{-1}^{-1/4}$ is the gap width \citep{2016PASJ...68...43K}. This is longer than the Kepler timescale in the range of our interest. If the gap width is smaller than the cavity size, i.e., $R_{\rm gap}<r_{\rm cav}$, we set $\rho_{\rm CSB}=\rho_{\rm AGN}$.

Because a smaller value of $v_{\rm kick}$ provides a higher mass accretion rate, we set the kick velocity to be $v_{\rm kick}=v_{\rm kick,cr}$. Then, the necessary conditions, $r_{\rm BHL}<r_{\rm Hill}$ and $r_{\rm BHL}<H$, are rewritten as $r_{\rm cav}>2.9\times10^{12}q_{-6}^{2/3}R_{16}$ cm and $r_{\rm cav}>6.3\times10^{12}q_{-6}\mathcal{H}_{-2.5}^{-1}R_{16}$ cm, respectively. We focus on the parameter space that satisfies these conditions, which is likely for most of the mergers. The kicked BH crosses the cavity in a timescale of 
\begin{equation}
t_{\rm cro}\approx r_{\rm cav}/v_{\rm kick}\simeq12~r_{\rm cav,13.5}v_{\rm kick,7.5}^{-1}\rm~day, 
\end{equation}
which dominates the time delay between the merger event and the EM transient. 

One can write $\dot{M}_{\rm BHL}\propto \rho_{\rm CSB}M_{\rm CSB}^2v_{\rm kick}^{-3}$. Noting $\rho_{\rm CSB}\propto\mathcal{H}^{-3}$ without a gap, the mass accretion rate is high for a small value of $\mathcal{H}$. A small value of $\mathcal{H}$ leads to a gap formation, in which the density depends on the aspect ratio as  $\rho\propto\mathcal{H}^2$. Then, the mass accretion rate is lower as $\mathcal{H}$ is smaller. Thus, the aspect ratio that makes $\chi_{\rm gap}\sim1$ provides the most efficient mass accretion onto the kicked BH, which would lead to the most luminous outflow-driven transients. We focus on such the most optimistic situation in the next subsection.

\subsection{Breakout emission from outflow bubbles}
\begin{table*}
\begin{center}
\begin{tabular}{|c|ccc|ccccccc|}
\hline
Model & $M_{\bullet}$ & $\mathcal{H}$ & $r_{\rm cav}$  & $v_{\rm kick}$ & $t_{\rm cro}$ & $t_{\rm bub}$ & $t_{\rm BBO}$ & $T_{\rm BBO}$ & $\log({\mathcal E}_{\rm BBO})$ & $\log(L_{\rm BBO})$ \\
 & ($M_\odot$) & & ($10^{13}$ cm) & (100 km s$^{-1}$) & (day) & (hour) & (min) & ($10^5$ K) & (erg) & ($\rm erg~s^{-1}$)\\
\hline
A & $10^8$ & 0.0020 & 4.0 & 4.9 & 9.4 & 11.2 & 8.0 & 4.7 & 47.3 & 44.6\\
B & $10^7$ & 0.0050 & 1.0 & 7.8 & 1.5 & 1.8 & 5.0 & 5.2 & 46.3 & 43.8 \\
\hline
\end{tabular}
\caption{Model parameters and resulting quantities for bubble breakout emission. Other parameters are $M_{\rm CSB}=150~M_\odot$, $\mathcal{R}=1.0\times10^3$, $\dot{m}=3.0$, $\alpha=0.1$, $\eta_w=0.32$, $v_w=10^{10}\rm~cm~s^{-1}$, $d_L=460$ Mpc ($z=0.1$). 
}\label{tab:param}
\end{center} 
\end{table*}

\begin{figure}
\begin{center}
\includegraphics[width=\linewidth]{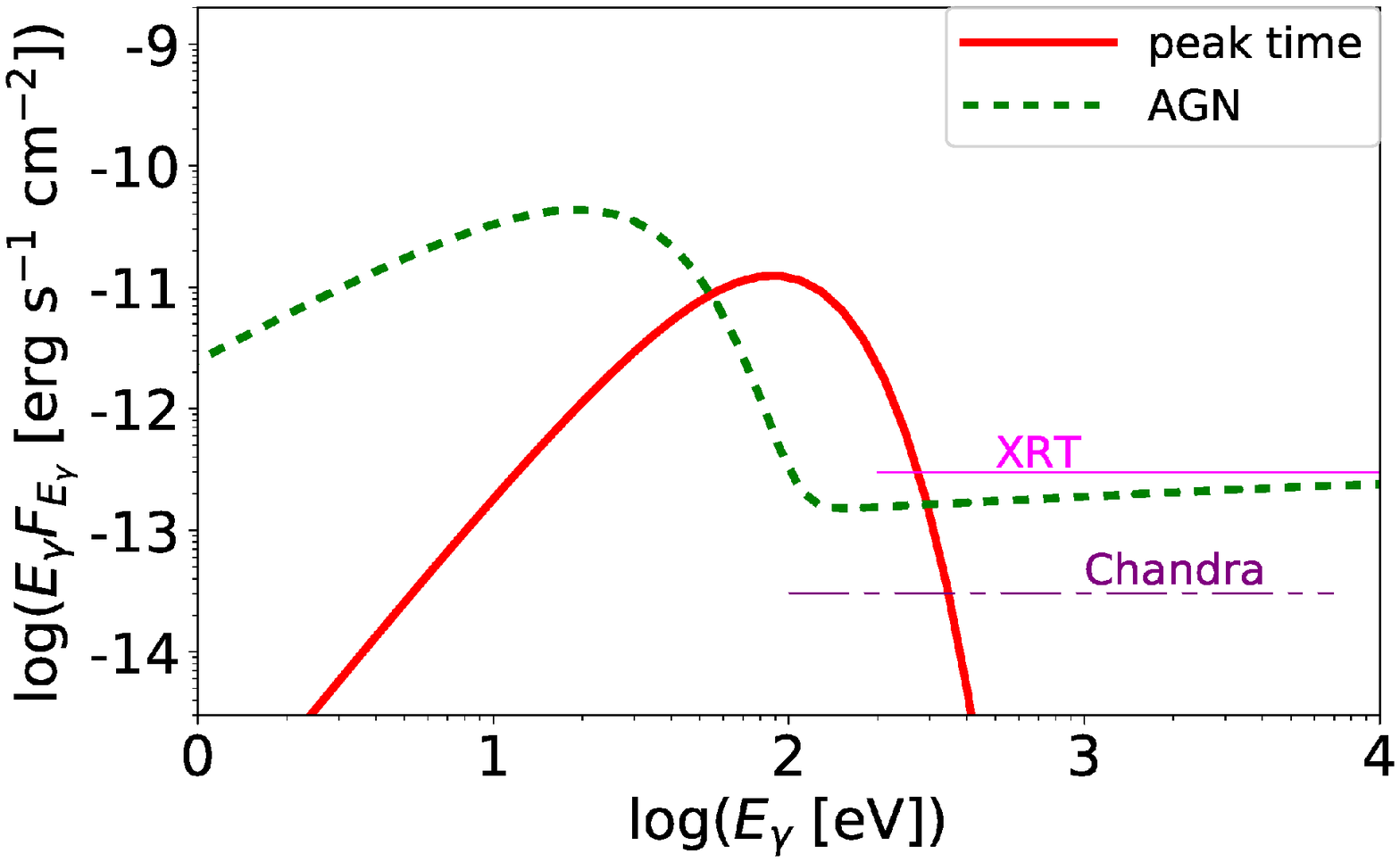}
\includegraphics[width=\linewidth]{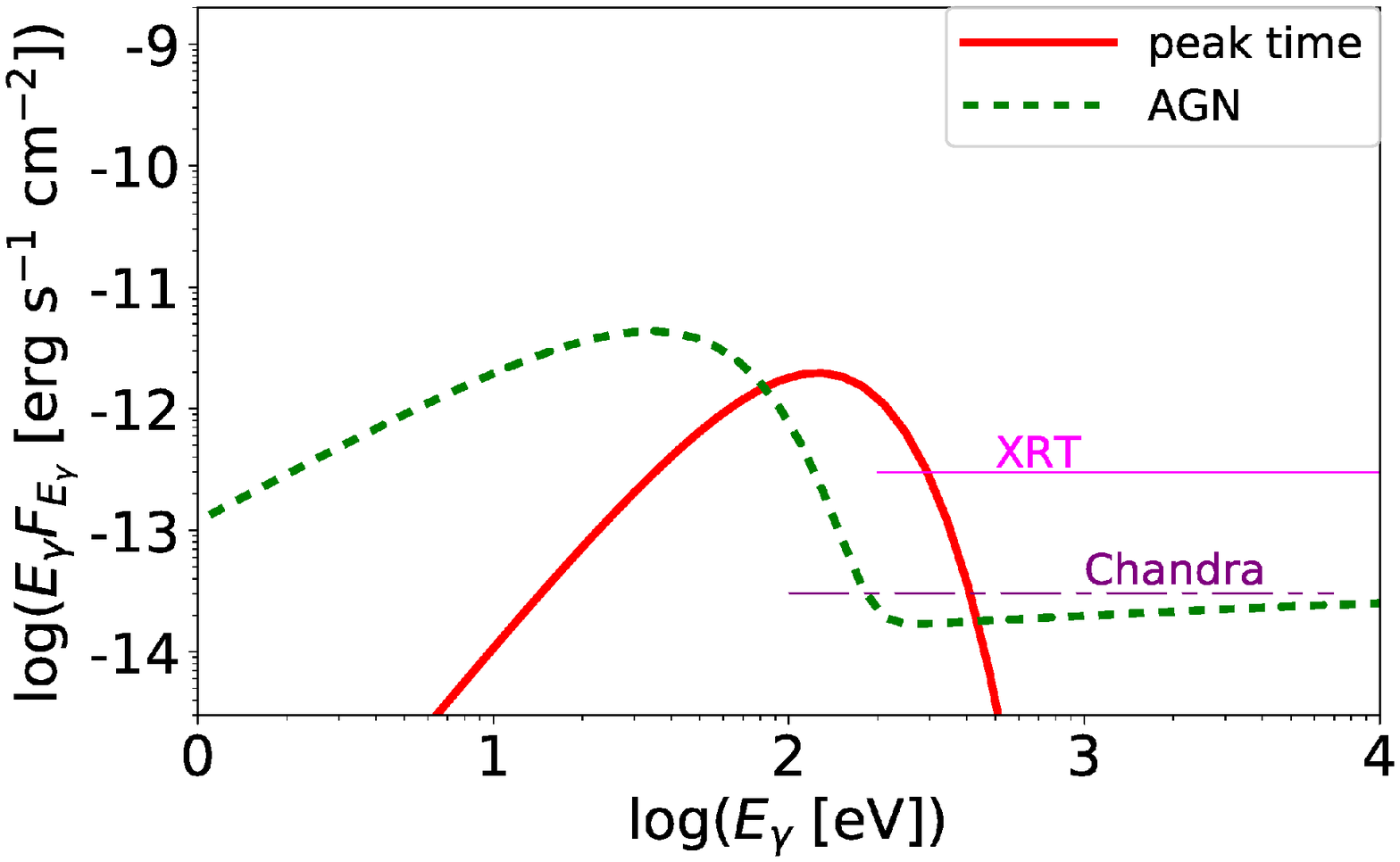}
\caption{{\bf Predicted spectra from the shock breakout by an outflow bubble from a merged BH for models A (top) and B (bottom).}
See Table~\ref{tab:param} for parameters. The AGN components are shown by green lines \added{(see text for details of the AGN components)}. The sensitivities to a $10^3$-sec transient for Swift XRT \citep{SwiftXRT05a} and Chandra \citep{2017MNRAS.467.4841B} are also indicated as thin-solid and thin-dotted-dashed lines. The outflow breakout emissions can be detectable with these current facilities.
\label{fig:obs}}
 \end{center}
\end{figure}

Because the accretion rate onto the kicked BH is super-Eddington, radiation-driven outflows are produced. The dynamics of the bubble expansion is similar to that discussed in Section~\ref{sec:cavity}, and thus the outflow bubble will break out from the disk in the timescale of 
\begin{equation}
t_{\rm bub}\approx\left(\frac{\rho_{\rm CSB} H^5}{0.53L_w}\right)^{1/3}\simeq7.4~\rho_{\rm CSB,-9}^{1/3}H_{13.5}^{5/3}L_{w,45.5}^{-1/3}\rm~hr. 
\end{equation}
This is much shorter than the Kepler time, and hence the outflow bubble breaks out from the AGN disk.

The photons inside the bubble start to diffuse out from the AGN disk at the time of the bubble breakout, namely, the photon diffusion time, $t_{\rm diff}\approx\Delta^2\kappa\rho_{\rm CSB}/c $, becomes equal to the bubble expansion time, $t_{\rm dyn}\approx\Delta/V_{\rm bub}$, where $\Delta$ is the thickness of the AGN disk above the bubble, $\kappa$ is the opacity for thermal photons, and $V_{\rm bub}\approx 3H/(5t_{\rm bub})\simeq 1.9\times10^9~H_{13.5}t_{\rm bub,4}^{-1}\rm~cm~s^{-1}$ is the bubble velocity at the time of the breakout. From this condition, we obtain $\Delta\approx c/(V_{\rm bub} \kappa \rho_{\rm CSB})$, and the duration of the bubble breakout emission is $t_{\rm BBO}=t_{\rm diff}=t_{\rm dyn}\approx c/(V_{\rm bub}^2\kappa\rho_{\rm CSB})\simeq75~V_{\rm bub,9}^{-2}\rho_{\rm CSB,-9}^{-1}$~s. Here, we use the electron scattering opacity, $\kappa=\sigma_T/m_p\simeq0.40$, for simplicity. From the shock jump condition, the temperature of the breakout photons is estimated to be $T_{\rm BBO}\approx(9\rho_{\rm CSB} V_{\rm bub}^2/4a_{\rm rad})^{1/4}\simeq7.4\times10^5\rho_{\rm CSB,-9}^{1/4}V_{s,9}^{1/2}{\rm~K}$, where $a_{\rm rad}$ is the radiation constant. The total energy of the breakout photons can be estimated to be ${\mathcal E}_{\rm BBO}\approx \pi H^2 \Delta a_{\rm rad}T_{\rm BBO}^4$. This energy is released in $t_{\rm BBO}$, so we can write the luminosity of the bubble breakout event as 
\begin{eqnarray}
L_{\rm BBO}&\approx&\frac{{\mathcal E}_{\rm BBO}}{t_{\rm BBO}}\approx\frac{9\pi H^2\rho_{\rm CSB}V_s^3}{4}\\
&\simeq&7.1\times10^{45}~\rho_{\rm CSB,-9}H_{13.5}^2V_{s,9}^{3}\rm~erg~s^{-1}.\nonumber
\end{eqnarray}
The breakout luminosity does not depend on $\kappa$, although it affects the duration of the breakout emission. In reality, the opacity may be higher due to the free-free absorption. This results in a longer transient, which may make the detection easier.

We focus on the detectability around the breakout time, where the emission peaks in the soft X-ray band. The emission peak lies in the UV band later, 
but the UV emission is easily outshone by the AGN disk emission. Let us compare the photon luminosity of the breakout emission to emission from the host AGN. Since the temperature of the breakout emission lies in the soft X-ray range, we construct the AGN spectrum in the UV and X-ray ranges. Here, we consider the multi-temperature black body emission from an optically thick disk \citep{pri81} for the UV emission and the Comptonized photons from a hot corona \citep{2018MNRAS.480.1819R} for the soft X-ray emission. For the AGN disk component, we consider an accretion flows onto Schwarzschild black hole, and use the radiation efficiency of $\eta_{\rm rad}\simeq0.06$. Then, the disk luminosity is estimated to be $L_{\rm disk}=\int L_{E_\gamma}dE_\gamma=\eta_{\rm rad}\dot{m}_{\rm AGN}L_{\rm Edd,AGN}$, where $L_{E_\gamma}$ is the differential luminosity. For the coronal component, we consider a power-law photon spectrum with an exponential cutoff, whose power-law index and cutoff energy are determined by the Eddington ratio, $\eta_{\rm rad}\dot{m}_{\rm AGN}$ \citep{2018MNRAS.480.1819R,Murase:2019vdl}. We normalize the X-ray luminosity using the bolometric correction of $\kappa_X\sim50$, and $L_{\rm crn}=\int L_{E_\gamma}dE_\gamma = L_{\rm disk}/50$ \added{\citep[e.g.][]{2007ApJ...654..731H}}.

Fig.~\ref{fig:obs} plots the resulting photon spectra for the outflow breakouts and the host AGN, whose parameters and resulting quantities are tabulated in Table~\ref{tab:param}. We see that the bubble breakout emission outshines the AGN emission in the soft X-ray range for both models at the peak time. This luminosity is above the sensitivity of current X-ray satellites, such as {\it Swift}-XRT and Chandra for $d_L\sim500$ Mpc. XMM-{\it Newton} also has a similar sensitivity and threshold energy to those for Chandra. 
The delay time of the transient to the merger event is equal to $t_{\rm cro}+t_{\rm bub}$, which is about a week (day) for model A (B). The typical timescale of the breakout emission, $t_{\rm BBO}$, is several minutes, corresponding to the rising time scale. 
The duration of the detectable EM emission can be several times longer, but details would depend on the density profile above the disk \citep[see][for a review]{2017hsn..book..967W}.

Our scenario is unlikely to be able to explain the optical counterpart of GW 190521. For the parameter set estimated by \cite{2020PhRvL.124y1102G}, an outflow cavity is expected to be produced. Then, the bubble breakout emission can produce a soft X-ray counterpart based on our scenario, but an optical counterpart is not expected. 
However, an optical transient can be produced if the X-rays are reprocessed by a dense material, such as AGN disk winds or broad-line clouds.

We mainly focused on sub-relativistic outflows launched by the disk around the merged BH. Given that the merged BH enters the disk region, it is also possible to have relativistic jets. 
\added{Relativistic jet formation in super-Edington systems is discussed in the context of jetted TDEs \citep{2011Sci...333..203B}, and the idea is supported by general relativistic radiation magnetohydrodynamic simulations \citep{2018ApJ...859L..20D}. Also,}
the formation of jets powered by the Bondi-Hoyle-Littleton accretion was discussed by \cite{Ioka:2016bil} in the context of Galactic EM counterparts of BBH merger remnants. In our scenario, the jet is launched once the merged BH enters the AGN disk. The jet is faster than the wind-driven bubble, and the bubble is dominated by a jet-induced cocoon as long as the jet luminosity is larger than the wind luminosity. With the jet head velocity $V_h=\beta_hc$ (that depends on the jet luminosity, density and position), it will breakout in $t_{\rm jbo}\sim H/V_h\sim10^4~H_{13.5}\beta_{h,-1}^{-1}$~s given that the jet direction is perpendicular to the AGN disk plane.  
Resulting cocoon emission can radiate a fraction of the energy with $L_jt_{\rm jbo}\sim10^{51}L_{j,47}t_{\rm jbo,4}$~erg, which could lead to an optical or UV transient.
Emission from the jet is brighter but the rate density of on-axis events is lower by the beaming factor.

\subsection{Event rates of BBH mergers with EM counterparts}
The rate of BBH mergers in AGN disks is estimated to be $0.02-60\rm~Gpc^{-3}~yr^{-1}$ \citep{2020ApJ...898...25T}. 
The event rate of the bubble breakout emission depends on many unknown parameters, such as distributions of the kick velocity and position of merger events. 
Thus, it is difficult to give quantitative estimates, but we here argue that the expected event rate of the outflow breakout emission is likely to be lower than the above rate of BBH mergers in AGN disks. 
Based on dedicated simulations in CSB mergers in AGN disks, the merger most likely occurs around 0.01 pc, which corresponds to $\mathcal{R}\sim10^3$ for $M_\bullet\sim10^8M_\odot$. Then, merger events with $v_{\rm kick}\gtrsim5\times10^2~\rm km~s^{-1}$ can go into the AGN disk again. The kick velocity may range from $10^2\rm~km~s^{-1}$ to $3\times10^3\rm~km~s^{-1}$, and let us assume a flat kick velocity distribution in linear space for simplicity. The luminosity of the breakout emission decreases with $L_{\rm BBO}\propto v_{\rm kick}^{-3}$, so a factor of a few higher kick velocity results in an order of magnitude dimmer event, which is readily outshone by AGN emission.  
Then, about 20\% of the kicked BHs have an appropriate kick velocity, i.e., $v_{\rm kick,cr} < v_{\rm kick} < 2v_{\rm kick,cr}$. For a non-spining BH, the kick direction is in the orbital plane, and the orbital plane of merging BBHs can be isotropically distributed \citep{2020ApJ...899...26T}. In this case, assuming a spherical cavity of size $H$, about 70\% of the kicked BHs have appropriate kick directions. Therefore, at most $\sim10$\% of the merged BH can go into the AGN disk again and produce breakout emission that may outshine the AGN emission.

\section{Summary and implications} \label{sec:summary}
In this work, we examined the mass accretion and outflow processes from CSBs embedded in AGN disks. Our conclusions are summarized below.
\begin{itemize}[leftmargin=*]
\item {\bf Compact binary mergers in AGN disks will mostly occur in cavities.} We showed that the accretion rate to a CSB is highly super-Eddington. This leads to a strong radiation-driven outflows from circum-binary disks, creating outflow bubbles inside the AGN disk. This bubble expands with time, and eventually breaks out from the AGN disk before the merger event in most of the parameter space.
\added{The outflows can be produced even from the progenitor of the CSB, i.e., a single compact object or a massive star. This means that the duration of the outflow production would be longer than our estimate in Section \ref{sec:cavity}. In this sense, our cavity formation condition is conservative and our argument is stronger.}

\item {\bf Detectable soft X-ray counterparts can be produced by recoiled remnant BHs entering the AGN disk.} 
If the merged BH is kicked toward the AGN disk with a high velocity, it gets into the unperturbed AGN disk again. This enables the BH to accrete gas from the disk at a super-Eddington rate. Then, a newly formed bubble is produced through the accretion process, which eventually breaks out from the AGN disk and causes a bright soft X-ray emission in days or weeks after the merger event. The duration of the breakout emission is about an hour. The luminosity of this emission can outshine the AGN disk in soft X-rays. This could be detectable with Swift-XRT or Chandra out to $d_L\sim500$\,Mpc.

Nevertheless, we expect that detecting outflow breakout emission will be challenging. First a bright outflow breakout requires an optimistic parameter set. A factor of a few higher kick velocity or higher aspect ratio results in the outflow transient dimmer than the AGN emission. Geometrically, about 70\% of the kicked BH can go into the AGN disk again if the kick direction distribution is isotropic. 
However, the threshold velocity required to go across the cavity depends on the location of the merger, which is highly uncertain. 
Further study is necessary to estimate the event rate more solidly.

For current instruments, successful follow-up observation will require the identification of the host galaxy in order to accommodate the relatively small fields of view of Swift-XRT and Chandra. Planned satellites, such as SVOM \citep{2016arXiv161006892W}, will have a wider field-of-view, which enables us to survey most of the error regions with a similar sensitivity to XRT. This greatly improves a chance to detect the outflow breakout emission. A lower threshold energy is also an important factor to detect the outflow breakout as the spectrum is very soft.

\item {\bf Reprocessed emission by the broadline region and molecular torus.}
A cavity is likely to be formed before the merger for a parameter set considered for GW190521 ($M_{\bullet}\sim10^8~M_\odot$, $\dot{m}\sim2.0$, $\mathcal{H}\sim0.01$, $\mathcal{R}\sim10^3$; \citealt{2020PhRvL.124y1102G}).
Breakout emission from an outflow bubble is possible for the above parameter set, but the outflow breakout emission does not explain the optical flux reported by \cite{2020PhRvL.124y1102G}. Thermal emission from disk-driven outflows is typically not as bright~\citep[e.g.,][]{2016ApJ...822L...9M,2017ApJ...851...53K}. Thus, explaining the optical flux require additional mechanisms such as the reprocessing by a dense AGN wind.

On the other hand, we point out that the soft X-ray emission can be reprocessed in the AGN environment \citep[e.g.,][]{Netzer:2015jna}. First, X-rays may be scattered by the broadline region, and $\sim10$\% of the X-ray flux may be scattered with a time scale from days to weeks. Second, the dusty torus is heated up and re-emits the X-ray energy in the IR band with a time scale from months to years. 

\item {\bf Outflows from CSBs could be persistent EM emitters long before the merger.} Since outflow bubbles exist before the merger events, the emission from the outflows may be observed as persistent or transient sources without GWs. As a long-lived signal produced by outflows, thermal emission from the outflows are expected. The luminosity is $\lesssim10^{42}\rm~erg~s^{-1}$ and the emission peaks at the optical and near UV bands, based on previous papers \citep[see, e.g.,][for transient cases]{KQ15a,2016ApJ...822L...9M,2017ApJ...851...52K,2017ApJ...851...53K}. At these bands, AGN emission is as bright as $10^{44}\rm~erg~s^{-1}$, which outshines the thermal emission. Resolving the two emission components requires facilities with a mili-arcsec resolution even for an AGN located at 10 Mpc from the Earth.

\item {\bf Short GRB jets from neutron star mergers are more likely to be seen without being choked, but gamma-ray spectra may be modified.} The mergers of BNS and NS-BH binaries are also expected in AGN disks, which can induce unique EM signals \citep[e.g.,][]{2020arXiv201108428Z,2021ApJ...906L...7P}. 
BNS and NS-BH mergers are expected to occur at the migration trap located in $\mathcal{R}<10^3$ \citep{2016ApJ...819L..17B} or $10^{-2}$ pc (corresponding to $\mathcal{R}\sim10^4$ for $M_\bullet=10^7M_\odot$) suggested by dedicated numerical calculations \citep{2020ApJ...898...25T}. As shown in Fig.~\ref{fig:cavity_param}, cavities are also formed for a BNS in a typical AGN at the expected radius. 

Because a cavity is still filled with an optically thick outflows, we will observe GRBs and kilonovae after the jet or ejecta penetrates the outflow component inside the cavity. However, the mass of the cavity is so low that the short GRB jet is not decelerated by the outflows below the photosphere. 
The photospheric radius of the outflows is estimated to be $r_{\rm ph}\approx \dot{M}_w\sigma_T/(4\pi v_wm_p)\sim3.2\times10^{11}~\dot{M}_{\rm CSB,22.5}v_{w,9}^{-1}\eta_{w,-0.5}\rm~cm$. The mass of the gas filling the cavity within the photosphere is then given by $M_{\rm cav}\sim 4\pi\rho_{\rm cav}r_{\rm ph}^3/3\sim1\times10^{24}\dot{M}_{\rm CSB,22.5}^2\eta_{w,-0.5}^2v_{w,9}^{-2}$ g, where $\rho_{\rm cav}\approx \dot{M}_{\rm CSB}\eta_w/(4\pi r_{\rm ph}^2 v_w)$ (see the bottom panel of Figure \ref{fig:mdot} for $\dot{M}_{\rm CSB}$ for a BNS).
The mass of the jet is estimated to be $M_{j,\rm iso}\sim E_{j,\rm iso}/(\Gamma_jc^2)\sim1\times10^{29}E_{j,\rm iso,52}\Gamma_{j,2}^{-1}$ g, where $E_{j,\rm iso}$ and $M_{j,\rm iso}$ are the isotropic equivalent energy and mass of the jets, respectively. Thus, the time lag between the GW and gamma-rays should be the same with that for a usual short GRB. The jets may be decelerated by the outflows below the internal dissipation radius, which makes a gamma-ray peak energy lower. Such a relatively low peak energy may be an indication of short GRBs occurred at outflow cavities in AGN disks. 
\end{itemize}


\acknowledgements
We thank Hiromichi Tagawa, Hidekazu Tanaka, and Yuki Tanaka for useful discussions. We also thank Zoltan Haiman for helpful comments. This work is supported by NSF Grant No.~AST-1908689 and JSPS KAKENHI No.~20H01901 and No.~20H05852 (K.M.), JSPS Research Fellowship and JSPS KAKENHI No.~19J00198 (S.S.K.). I.B. acknowledges support from the Alfred P. Sloan Foundation and the University of Florida.

\bibliographystyle{aasjournal}
\bibliography{ssk,kmurase}

\hyphenation{Post-Script Sprin-ger}
\begin{thebibliography}{}
\expandafter\ifx\csname natexlab\endcsname\relax\def\natexlab#1{#1}\fi
\providecommand{\url}[1]{\href{#1}{#1}}
\providecommand{\dodoi}[1]{doi:~\href{http://doi.org/#1}{\nolinkurl{#1}}}
\providecommand{\doeprint}[1]{\href{http://ascl.net/#1}{\nolinkurl{http://ascl.net/#1}}}
\providecommand{\doarXiv}[1]{\href{https://arxiv.org/abs/#1}{\nolinkurl{https://arxiv.org/abs/#1}}}

\bibitem[{{Aasi} {et~al.}(2015)}]{2015CQGra..32g4001L}
{Aasi}, J., {et~al.} 2015, Class Quantum Gravity, 32, 074001,
  \dodoi{10.1088/0264-9381/32/7/074001}

\bibitem[{Abbott {et~al.}(2020{\natexlab{a}})}]{Abbott:2020gyp}
Abbott, R., {et~al.} 2020{\natexlab{a}}.
\newblock \doarXiv{2010.14533}

\bibitem[{Abbott {et~al.}(2020{\natexlab{b}})}]{Abbott:2020niy}
---. 2020{\natexlab{b}}.
\newblock \doarXiv{2010.14527}

\bibitem[{Abbott {et~al.}(2020{\natexlab{c}})}]{LIGOScientific:2020stg}
---. 2020{\natexlab{c}}, Phys. Rev. D, 102, 043015,
  \dodoi{10.1103/PhysRevD.102.043015}

\bibitem[{Abbott {et~al.}(2020{\natexlab{d}})}]{Abbott:2020khf}
---. 2020{\natexlab{d}}, Astrophys. J. Lett., 896, L44,
  \dodoi{10.3847/2041-8213/ab960f}

\bibitem[{Abbott {et~al.}(2020{\natexlab{e}})}]{Abbott:2020tfl}
---. 2020{\natexlab{e}}, Phys. Rev. Lett., 125, 101102,
  \dodoi{10.1103/PhysRevLett.125.101102}

\bibitem[{Abbott {et~al.}(2020{\natexlab{f}})}]{Abbott:2020mjq}
---. 2020{\natexlab{f}}, Astrophys. J. Lett., 900, L13,
  \dodoi{10.3847/2041-8213/aba493}

\bibitem[{{Acernese} {et~al.}(2015)}]{2015CQGra..32b4001A}
{Acernese}, F., {et~al.} 2015, Class Quantum Gravity, 32, 024001,
  \dodoi{10.1088/0264-9381/32/2/024001}

\bibitem[{{Bartos} {et~al.}(2017){Bartos}, {Kocsis}, {Haiman}, \&
  {M{\'a}rka}}]{BKH17a}
{Bartos}, I., {Kocsis}, B., {Haiman}, Z., \& {M{\'a}rka}, S. 2017, \apj, 835,
  165, \dodoi{10.3847/1538-4357/835/2/165}

\bibitem[{{Baruteau} {et~al.}(2011){Baruteau}, {Meru}, \&
  {Paardekooper}}]{2011MNRAS.416.1971B}
{Baruteau}, C., {Meru}, F., \& {Paardekooper}, S.-J. 2011, \mnras, 416, 1971,
  \dodoi{10.1111/j.1365-2966.2011.19172.x}

\bibitem[{{Bauer} {et~al.}(2017){Bauer}, {Treister}, {Schawinski}, {Schulze},
  {Luo}, {Alexander}, {Brandt}, {Comastri}, {Forster}, {Gilli}, {Kann},
  {Maeda}, {Nomoto}, {Paolillo}, {Ranalli}, {Schneider}, {Shemmer}, {Tanaka},
  {Tolstov}, {Tominaga}, {Tozzi}, {Vignali}, {Wang}, {Xue}, \&
  {Yang}}]{2017MNRAS.467.4841B}
{Bauer}, F.~E., {Treister}, E., {Schawinski}, K., {et~al.} 2017, \mnras, 467,
  4841, \dodoi{10.1093/mnras/stx417}

\bibitem[{{Belczynski}(2020)}]{2020arXiv200913526B}
{Belczynski}, K. 2020, arXiv e-prints, arXiv:2009.13526.
\newblock \doarXiv{2009.13526}

\bibitem[{{Belczynski} {et~al.}(2016){Belczynski}, {Holz}, {Bulik}, \&
  {O'Shaughnessy}}]{BHB16a}
{Belczynski}, K., {Holz}, D.~E., {Bulik}, T., \& {O'Shaughnessy}, R. 2016,
  \nat, 534, 512, \dodoi{10.1038/nature18322}

\bibitem[{{Bellovary} {et~al.}(2016){Bellovary}, {Mac Low}, {McKernan}, \&
  {Ford}}]{2016ApJ...819L..17B}
{Bellovary}, J.~M., {Mac Low}, M.-M., {McKernan}, B., \& {Ford}, K.~E.~S. 2016,
  \apjl, 819, L17, \dodoi{10.3847/2041-8205/819/2/L17}

\bibitem[{{Bloom} {et~al.}(2011){Bloom}, {Giannios}, {Metzger}, {Cenko},
  {Perley}, {Butler}, {Tanvir}, {Levan}, {O'Brien}, {Strubbe}, {De Colle},
  {Ramirez-Ruiz}, {Lee}, {Nayakshin}, {Quataert}, {King}, {Cucchiara},
  {Guillochon}, {Bower}, {Fruchter}, {Morgan}, \& {van der
  Horst}}]{2011Sci...333..203B}
{Bloom}, J.~S., {Giannios}, D., {Metzger}, B.~D., {et~al.} 2011, Science, 333,
  203, \dodoi{10.1126/science.1207150}

\bibitem[{{Bondi}(1952)}]{Bon52a}
{Bondi}, H. 1952, \mnras, 112, 195, \dodoi{10.1093/mnras/112.2.195}

\bibitem[{{Burrows} {et~al.}(2005){Burrows}, {Hill}, {Nousek}, {Kennea},
  {Wells}, {Osborne}, {Abbey}, {Beardmore}, {Mukerjee}, {Short}, {Chincarini},
  {Campana}, {Citterio}, {Moretti}, {Pagani}, {Tagliaferri}, {Giommi},
  {Capalbi}, {Tamburelli}, {Angelini}, {Cusumano}, {Br{\"a}uninger}, {Burkert},
  \& {Hartner}}]{SwiftXRT05a}
{Burrows}, D.~N., {Hill}, J.~E., {Nousek}, J.~A., {et~al.} 2005, \ssr, 120,
  165, \dodoi{10.1007/s11214-005-5097-2}

\bibitem[{{Campanelli} {et~al.}(2007{\natexlab{a}}){Campanelli}, {Lousto},
  {Zlochower}, \& {Merritt}}]{2007ApJ...659L...5C}
{Campanelli}, M., {Lousto}, C., {Zlochower}, Y., \& {Merritt}, D.
  2007{\natexlab{a}}, \apjl, 659, L5, \dodoi{10.1086/516712}

\bibitem[{{Campanelli} {et~al.}(2007{\natexlab{b}}){Campanelli}, {Lousto},
  {Zlochower}, \& {Merritt}}]{2007PhRvL..98w1102C}
{Campanelli}, M., {Lousto}, C.~O., {Zlochower}, Y., \& {Merritt}, D.
  2007{\natexlab{b}}, \prl, 98, 231102, \dodoi{10.1103/PhysRevLett.98.231102}

\bibitem[{{Centrella} {et~al.}(2010){Centrella}, {Baker}, {Kelly}, \& {van
  Meter}}]{2010RvMP...82.3069C}
{Centrella}, J., {Baker}, J.~G., {Kelly}, B.~J., \& {van Meter}, J.~R. 2010,
  Reviews of Modern Physics, 82, 3069, \dodoi{10.1103/RevModPhys.82.3069}

\bibitem[{{Costa} {et~al.}(2020){Costa}, {Bressan}, {Mapelli}, {Marigo},
  {Iorio}, \& {Spera}}]{2020arXiv201002242C}
{Costa}, G., {Bressan}, A., {Mapelli}, M., {et~al.} 2020, arXiv e-prints,
  arXiv:2010.02242.
\newblock \doarXiv{2010.02242}

\bibitem[{{Dai} {et~al.}(2018){Dai}, {McKinney}, {Roth}, {Ramirez-Ruiz}, \&
  {Miller}}]{2018ApJ...859L..20D}
{Dai}, L., {McKinney}, J.~C., {Roth}, N., {Ramirez-Ruiz}, E., \& {Miller},
  M.~C. 2018, \apjl, 859, L20, \dodoi{10.3847/2041-8213/aab429}

\bibitem[{{D'Angelo} {et~al.}(2003){D'Angelo}, {Henning}, \&
  {Kley}}]{2003ApJ...599..548D}
{D'Angelo}, G., {Henning}, T., \& {Kley}, W. 2003, \apj, 599, 548,
  \dodoi{10.1086/379224}

\bibitem[{{de Mink} \& {King}(2017)}]{MK17a}
{de Mink}, S.~E., \& {King}, A. 2017, \apjl, 839, L7,
  \dodoi{10.3847/2041-8213/aa67f3}

\bibitem[{{D'Orazio} {et~al.}(2013){D'Orazio}, {Haiman}, \&
  {MacFadyen}}]{2013MNRAS.436.2997D}
{D'Orazio}, D.~J., {Haiman}, Z., \& {MacFadyen}, A. 2013, \mnras, 436, 2997,
  \dodoi{10.1093/mnras/stt1787}

\bibitem[{{Edgar}(2004)}]{Edg04a}
{Edgar}, R. 2004, New Astron. Rev., 48, 843,
  \dodoi{10.1016/j.newar.2004.06.001}

\bibitem[{{Farr} {et~al.}(2011){Farr}, {Sravan}, {Cantrell}, {Kreidberg},
  {Bailyn}, {Mandel}, \& {Kalogera}}]{2011ApJ...741..103F}
{Farr}, W.~M., {Sravan}, N., {Cantrell}, A., {et~al.} 2011, \apj, 741, 103,
  \dodoi{10.1088/0004-637X/741/2/103}

\bibitem[{{Farrell} {et~al.}(2020){Farrell}, {Groh}, {Hirschi}, {Murphy},
  {Kaiser}, {Ekstr{\"o}m}, {Georgy}, \& {Meynet}}]{2020arXiv200906585F}
{Farrell}, E.~J., {Groh}, J.~H., {Hirschi}, R., {et~al.} 2020, arXiv e-prints,
  arXiv:2009.06585.
\newblock \doarXiv{2009.06585}

\bibitem[{{Farris} {et~al.}(2014){Farris}, {Duffell}, {MacFadyen}, \&
  {Haiman}}]{FDM14a}
{Farris}, B.~D., {Duffell}, P., {MacFadyen}, A.~I., \& {Haiman}, Z. 2014, \apj,
  783, 134, \dodoi{10.1088/0004-637X/783/2/134}

\bibitem[{{Fujii} {et~al.}(2017){Fujii}, {Tanikawa}, \& {Makino}}]{FTM17a}
{Fujii}, M., {Tanikawa}, A., \& {Makino}, J. 2017, ArXiv e-prints.
\newblock \doarXiv{1709.02058}

\bibitem[{{Gayathri} {et~al.}(2020{\natexlab{a}}){Gayathri}, {Bartos},
  {Haiman}, {Klimenko}, {Kocsis}, {M{\'a}rka}, \& {Yang}}]{2020ApJ...890L..20G}
{Gayathri}, V., {Bartos}, I., {Haiman}, Z., {et~al.} 2020{\natexlab{a}}, \apjl,
  890, L20, \dodoi{10.3847/2041-8213/ab745d}

\bibitem[{{Gayathri} {et~al.}(2020{\natexlab{b}}){Gayathri}, {Healy}, {Lange},
  {O'Brien}, {Szczepanczyk}, {Bartos}, {Campanelli}, {Klimenko}, {Lousto}, \&
  {O'Shaughnessy}}]{2020arXiv200905461G}
{Gayathri}, V., {Healy}, J., {Lange}, J., {et~al.} 2020{\natexlab{b}}, arXiv
  e-prints, arXiv:2009.05461.
\newblock \doarXiv{2009.05461}

\bibitem[{{Gonz{\'a}lez} {et~al.}(2007){Gonz{\'a}lez}, {Sperhake},
  {Br{\"u}gmann}, {Hannam}, \& {Husa}}]{2007PhRvL..98i1101G}
{Gonz{\'a}lez}, J.~A., {Sperhake}, U., {Br{\"u}gmann}, B., {Hannam}, M., \&
  {Husa}, S. 2007, \prl, 98, 091101, \dodoi{10.1103/PhysRevLett.98.091101}

\bibitem[{{Graham} {et~al.}(2020){Graham}, {Ford}, {McKernan}, {Ross}, {Stern},
  {Burdge}, {Coughlin}, {Djorgovski}, {Drake}, {Duev}, {Kasliwal}, {Mahabal},
  {van Velzen}, {Belecki}, {Bellm}, {Burruss}, {Cenko}, {Cunningham}, {Helou},
  {Kulkarni}, {Masci}, {Prince}, {Reiley}, {Rodriguez}, {Rusholme}, {Smith}, \&
  {Soumagnac}}]{2020PhRvL.124y1102G}
{Graham}, M.~J., {Ford}, K.~E.~S., {McKernan}, B., {et~al.} 2020, \prl, 124,
  251102, \dodoi{10.1103/PhysRevLett.124.251102}

\bibitem[{{Herrmann} {et~al.}(2007){Herrmann}, {Hinder}, {Shoemaker}, \&
  {Laguna}}]{2007CQGra..24S..33H}
{Herrmann}, F., {Hinder}, I., {Shoemaker}, D., \& {Laguna}, P. 2007, Classical
  and Quantum Gravity, 24, S33, \dodoi{10.1088/0264-9381/24/12/S04}

\bibitem[{{Hopkins} {et~al.}(2007){Hopkins}, {Richards}, \&
  {Hernquist}}]{2007ApJ...654..731H}
{Hopkins}, P.~F., {Richards}, G.~T., \& {Hernquist}, L. 2007, \apj, 654, 731,
  \dodoi{10.1086/509629}

\bibitem[{{Hoyle} \& {Lyttleton}(1939)}]{HL39a}
{Hoyle}, F., \& {Lyttleton}, R.~A. 1939, Proceedings of the Cambridge
  Philosophical Society, 35, 405, \dodoi{10.1017/S0305004100021150}

\bibitem[{Ioka {et~al.}(2017)Ioka, Matsumoto, Teraki, Kashiyama, \&
  Murase}]{Ioka:2016bil}
Ioka, K., Matsumoto, T., Teraki, Y., Kashiyama, K., \& Murase, K. 2017, Mon.
  Not. Roy. Astron. Soc., 470, 3332, \dodoi{10.1093/mnras/stx1337}

\bibitem[{{Jiang} {et~al.}(2014){Jiang}, {Stone}, \& {Davis}}]{JSD14a}
{Jiang}, Y.-F., {Stone}, J.~M., \& {Davis}, S.~W. 2014, \apj, 796, 106,
  \dodoi{10.1088/0004-637X/796/2/106}

\bibitem[{{Jiao} {et~al.}(2015){Jiao}, {Mineshige}, {Takeuchi}, \&
  {Ohsuga}}]{2015ApJ...806...93J}
{Jiao}, C.-L., {Mineshige}, S., {Takeuchi}, S., \& {Ohsuga}, K. 2015, \apj,
  806, 93, \dodoi{10.1088/0004-637X/806/1/93}

\bibitem[{{Kanagawa} {et~al.}(2016){Kanagawa}, {Muto}, {Tanaka}, {Tanigawa},
  {Takeuchi}, {Tsukagoshi}, \& {Momose}}]{2016PASJ...68...43K}
{Kanagawa}, K.~D., {Muto}, T., {Tanaka}, H., {et~al.} 2016, \pasj, 68, 43,
  \dodoi{10.1093/pasj/psw037}

\bibitem[{{Kanagawa} {et~al.}(2015){Kanagawa}, {Tanaka}, {Muto}, {Tanigawa}, \&
  {Takeuchi}}]{2015MNRAS.448..994K}
{Kanagawa}, K.~D., {Tanaka}, H., {Muto}, T., {Tanigawa}, T., \& {Takeuchi}, T.
  2015, \mnras, 448, 994, \dodoi{10.1093/mnras/stv025}

\bibitem[{{Kanagawa} {et~al.}(2018){Kanagawa}, {Tanaka}, \&
  {Szuszkiewicz}}]{2018ApJ...861..140K}
{Kanagawa}, K.~D., {Tanaka}, H., \& {Szuszkiewicz}, E. 2018, \apj, 861, 140,
  \dodoi{10.3847/1538-4357/aac8d9}

\bibitem[{{Kashiyama} \& {Quataert}(2015)}]{KQ15a}
{Kashiyama}, K., \& {Quataert}, E. 2015, \mnras, 451, 2656,
  \dodoi{10.1093/mnras/stv1164}

\bibitem[{{Kato} {et~al.}(2008){Kato}, {Fukue}, \& {Mineshige}}]{KFM08a}
{Kato}, S., {Fukue}, J., \& {Mineshige}, S. 2008, {Black-Hole Accretion Disks
  --- Towards a New Paradigm ---} (Kyoto University Press)

\bibitem[{Kimura {et~al.}(2018)Kimura, Murase, Bartos, Ioka, Heng, \&
  M\'esz\'aros}]{Kimura:2018vvz}
Kimura, S.~S., Murase, K., Bartos, I., {et~al.} 2018, Phys.\ Rev.\ D, 98,
  043020, \dodoi{10.1103/PhysRevD.98.043020}

\bibitem[{{Kimura} {et~al.}(2017{\natexlab{a}}){Kimura}, {Murase}, \&
  {M{\'e}sz{\'a}ros}}]{2017ApJ...851...52K}
{Kimura}, S.~S., {Murase}, K., \& {M{\'e}sz{\'a}ros}, P. 2017{\natexlab{a}},
  \apj, 851, 52, \dodoi{10.3847/1538-4357/aa989b}

\bibitem[{{Kimura} {et~al.}(2017{\natexlab{b}}){Kimura}, {Murase}, \&
  {M{\'e}sz{\'a}ros}}]{2017ApJ...851...53K}
---. 2017{\natexlab{b}}, \apj, 851, 53, \dodoi{10.3847/1538-4357/aa988b}

\bibitem[{{Kinugawa} {et~al.}(2014){Kinugawa}, {Inayoshi}, {Hotokezaka},
  {Nakauchi}, \& {Nakamura}}]{KIH14a}
{Kinugawa}, T., {Inayoshi}, K., {Hotokezaka}, K., {Nakauchi}, D., \&
  {Nakamura}, T. 2014, \mnras, 442, 2963, \dodoi{10.1093/mnras/stu1022}

\bibitem[{{Kitaki} {et~al.}(2018){Kitaki}, {Mineshige}, {Ohsuga}, \&
  {Kawashima}}]{2018PASJ...70..108K}
{Kitaki}, T., {Mineshige}, S., {Ohsuga}, K., \& {Kawashima}, T. 2018, \pasj,
  70, 108, \dodoi{10.1093/pasj/psy110}

\bibitem[{{Koo} \& {McKee}(1992)}]{1992ApJ...388...93K}
{Koo}, B.-C., \& {McKee}, C.~F. 1992, \apj, 388, 93, \dodoi{10.1086/171132}

\bibitem[{{Li} {et~al.}(2021){Li}, {Dempsey}, {Li}, {Li}, \&
  {Li}}]{2021arXiv210109406L}
{Li}, Y.-P., {Dempsey}, A.~M., {Li}, S., {Li}, H., \& {Li}, J. 2021, arXiv
  e-prints, arXiv:2101.09406.
\newblock \doarXiv{2101.09406}

\bibitem[{{Li} {et~al.}(2011){Li}, {Ho}, \& {Wang}}]{lhw11}
{Li}, Y.-R., {Ho}, L.~C., \& {Wang}, J.-M. 2011, \apj, 742, 33,
  \dodoi{10.1088/0004-637X/742/1/33}

\bibitem[{{Lippai} {et~al.}(2008){Lippai}, {Frei}, \&
  {Haiman}}]{2008ApJ...676L...5L}
{Lippai}, Z., {Frei}, Z., \& {Haiman}, Z. 2008, \apjl, 676, L5,
  \dodoi{10.1086/587034}

\bibitem[{{Liu} \& {Bromm}(2020)}]{2020ApJ...903L..40L}
{Liu}, B., \& {Bromm}, V. 2020, \apjl, 903, L40,
  \dodoi{10.3847/2041-8213/abc552}

\bibitem[{{Lubow} {et~al.}(1999){Lubow}, {Seibert}, \&
  {Artymowicz}}]{1999ApJ...526.1001L}
{Lubow}, S.~H., {Seibert}, M., \& {Artymowicz}, P. 1999, \apj, 526, 1001,
  \dodoi{10.1086/308045}

\bibitem[{{Machida} {et~al.}(2010){Machida}, {Kokubo}, {Inutsuka}, \&
  {Matsumoto}}]{2010MNRAS.405.1227M}
{Machida}, M.~N., {Kokubo}, E., {Inutsuka}, S.-I., \& {Matsumoto}, T. 2010,
  \mnras, 405, 1227, \dodoi{10.1111/j.1365-2966.2010.16527.x}

\bibitem[{{McKernan} {et~al.}(2012){McKernan}, {Ford}, {Lyra}, \&
  {Perets}}]{2012MNRAS.425..460M}
{McKernan}, B., {Ford}, K.~E.~S., {Lyra}, W., \& {Perets}, H.~B. 2012, \mnras,
  425, 460, \dodoi{10.1111/j.1365-2966.2012.21486.x}

\bibitem[{McKernan {et~al.}(2019)McKernan, Ford, Bartos, Graham, Lyra, Marka,
  Marka, Ross, Stern, \& Yang}]{McKernan_2019}
McKernan, B., Ford, K. E.~S., Bartos, I., {et~al.} 2019, The Astrophysical
  Journal, 884, L50, \dodoi{10.3847/2041-8213/ab4886}

\bibitem[{{Milosavljevi{\'c}} {et~al.}(2009){Milosavljevi{\'c}}, {Bromm},
  {Couch}, \& {Oh}}]{2009ApJ...698..766M}
{Milosavljevi{\'c}}, M., {Bromm}, V., {Couch}, S.~M., \& {Oh}, S.~P. 2009,
  \apj, 698, 766, \dodoi{10.1088/0004-637X/698/1/766}

\bibitem[{{Moody} {et~al.}(2019){Moody}, {Shi}, \&
  {Stone}}]{2019ApJ...875...66M}
{Moody}, M. S.~L., {Shi}, J.-M., \& {Stone}, J.~M. 2019, \apj, 875, 66,
  \dodoi{10.3847/1538-4357/ab09ee}

\bibitem[{Murase \& Ioka(2013)}]{Murase:2013ffa}
Murase, K., \& Ioka, K. 2013, Phys.Rev.Lett., 111, 121102,
  \dodoi{10.1103/PhysRevLett.111.121102}

\bibitem[{{Murase} {et~al.}(2016){Murase}, {Kashiyama}, {M{\'e}sz{\'a}ros},
  {Shoemaker}, \& {Senno}}]{2016ApJ...822L...9M}
{Murase}, K., {Kashiyama}, K., {M{\'e}sz{\'a}ros}, P., {Shoemaker}, I., \&
  {Senno}, N. 2016, \apjl, 822, L9, \dodoi{10.3847/2041-8205/822/1/L9}

\bibitem[{Murase {et~al.}(2020)Murase, Kimura, \& Meszaros}]{Murase:2019vdl}
Murase, K., Kimura, S.~S., \& Meszaros, P. 2020, Phys. Rev. Lett., 125, 011101,
  \dodoi{10.1103/PhysRevLett.125.011101}

\bibitem[{Netzer(2015)}]{Netzer:2015jna}
Netzer, H. 2015, Ann. Rev. Astron. Astrophys., 53, 365,
  \dodoi{10.1146/annurev-astro-082214-122302}

\bibitem[{{Nixon} {et~al.}(2013){Nixon}, {King}, \&
  {Price}}]{2013MNRAS.434.1946N}
{Nixon}, C., {King}, A., \& {Price}, D. 2013, \mnras, 434, 1946,
  \dodoi{10.1093/mnras/stt1136}

\bibitem[{{Ohsuga} {et~al.}(2005){Ohsuga}, {Mori}, {Nakamoto}, \&
  {Mineshige}}]{OMN05a}
{Ohsuga}, K., {Mori}, M., {Nakamoto}, T., \& {Mineshige}, S. 2005, \apj, 628,
  368, \dodoi{10.1086/430728}

\bibitem[{{{\"O}zel} {et~al.}(2012){{\"O}zel}, {Psaltis}, {Narayan}, \& {Santos
  Villarreal}}]{2012ApJ...757...55O}
{{\"O}zel}, F., {Psaltis}, D., {Narayan}, R., \& {Santos Villarreal}, A. 2012,
  \apj, 757, 55, \dodoi{10.1088/0004-637X/757/1/55}

\bibitem[{{Perna} {et~al.}(2021){Perna}, {Lazzati}, \&
  {Cantiello}}]{2021ApJ...906L...7P}
{Perna}, R., {Lazzati}, D., \& {Cantiello}, M. 2021, \apjl, 906, L7,
  \dodoi{10.3847/2041-8213/abd319}

\bibitem[{{Pringle}(1981)}]{pri81}
{Pringle}, J.~E. 1981, \araa, 19, 137,
  \dodoi{10.1146/annurev.aa.19.090181.001033}

\bibitem[{{Proga}(2007)}]{2007ApJ...661..693P}
{Proga}, D. 2007, \apj, 661, 693, \dodoi{10.1086/515389}

\bibitem[{{Rezzolla} {et~al.}(2008){Rezzolla}, {Barausse}, {Dorband},
  {Pollney}, {Reisswig}, {Seiler}, \& {Husa}}]{2008PhRvD..78d4002R}
{Rezzolla}, L., {Barausse}, E., {Dorband}, E.~N., {et~al.} 2008, \prd, 78,
  044002, \dodoi{10.1103/PhysRevD.78.044002}

\bibitem[{{Ricci} {et~al.}(2018){Ricci}, {Ho}, {Fabian}, {Trakhtenbrot},
  {Koss}, {Ueda}, {Lohfink}, {Shimizu}, {Bauer}, {Mushotzky}, {Schawinski},
  {Paltani}, {Lamperti}, {Treister}, \& {Oh}}]{2018MNRAS.480.1819R}
{Ricci}, C., {Ho}, L.~C., {Fabian}, A.~C., {et~al.} 2018, \mnras, 480, 1819,
  \dodoi{10.1093/mnras/sty1879}

\bibitem[{{Rodriguez} {et~al.}(2016){Rodriguez}, {Haster}, {Chatterjee},
  {Kalogera}, \& {Rasio}}]{RHC16a}
{Rodriguez}, C.~L., {Haster}, C.-J., {Chatterjee}, S., {Kalogera}, V., \&
  {Rasio}, F.~A. 2016, \apjl, 824, L8, \dodoi{10.3847/2041-8205/824/1/L8}

\bibitem[{{Safarzadeh} \& {Haiman}(2020)}]{2020ApJ...903L..21S}
{Safarzadeh}, M., \& {Haiman}, Z. 2020, \apjl, 903, L21,
  \dodoi{10.3847/2041-8213/abc253}

\bibitem[{{Samsing} {et~al.}(2020){Samsing}, {Bartos}, {D'Orazio}, {Haiman},
  {Kocsis}, {Leigh}, {Liu}, {Pessah}, \& {Tagawa}}]{2020arXiv201009765S}
{Samsing}, J., {Bartos}, I., {D'Orazio}, D.~J., {et~al.} 2020, arXiv e-prints,
  arXiv:2010.09765.
\newblock \doarXiv{2010.09765}

\bibitem[{{S{\c a}dowski} {et~al.}(2014){S{\c a}dowski}, {Narayan}, {McKinney},
  \& {Tchekhovskoy}}]{SNM14a}
{S{\c a}dowski}, A., {Narayan}, R., {McKinney}, J.~C., \& {Tchekhovskoy}, A.
  2014, \mnras, 439, 503, \dodoi{10.1093/mnras/stt2479}

\bibitem[{{S{\c a}dowski} {et~al.}(2013){S{\c a}dowski}, {Narayan}, {Penna}, \&
  {Zhu}}]{SNP13a}
{S{\c a}dowski}, A., {Narayan}, R., {Penna}, R., \& {Zhu}, Y. 2013, \mnras,
  436, 3856, \dodoi{10.1093/mnras/stt1881}

\bibitem[{Senno {et~al.}(2016)Senno, Murase, \& M\'esz\'aros}]{Senno:2015tsn}
Senno, N., Murase, K., \& M\'esz\'aros, P. 2016, Phys. Rev., D93, 083003,
  \dodoi{10.1103/PhysRevD.93.083003}

\bibitem[{{Shakura} \& {Sunyaev}(1973)}]{ss73}
{Shakura}, N.~I., \& {Sunyaev}, R.~A. 1973, \aap, 24, 337

\bibitem[{{Shapiro} \& {Teukolsky}(1983)}]{ST83a}
{Shapiro}, S.~L., \& {Teukolsky}, S.~A. 1983, {Black holes, white dwarfs, and
  neutron stars: The physics of compact objects} (Wiley-VCH)

\bibitem[{{Stone} {et~al.}(2017){Stone}, {Metzger}, \& {Haiman}}]{SMH17a}
{Stone}, N.~C., {Metzger}, B.~D., \& {Haiman}, Z. 2017, \mnras, 464, 946,
  \dodoi{10.1093/mnras/stw2260}

\bibitem[{{Sugimura} {et~al.}(2017){Sugimura}, {Hosokawa}, {Yajima}, \&
  {Omukai}}]{2017MNRAS.469...62S}
{Sugimura}, K., {Hosokawa}, T., {Yajima}, H., \& {Omukai}, K. 2017, \mnras,
  469, 62, \dodoi{10.1093/mnras/stx769}

\bibitem[{{Tagawa} {et~al.}(2020{\natexlab{a}}){Tagawa}, {Haiman}, {Bartos}, \&
  {Kocsis}}]{2020ApJ...899...26T}
{Tagawa}, H., {Haiman}, Z., {Bartos}, I., \& {Kocsis}, B. 2020{\natexlab{a}},
  \apj, 899, 26, \dodoi{10.3847/1538-4357/aba2cc}

\bibitem[{{Tagawa} {et~al.}(2020{\natexlab{b}}){Tagawa}, {Haiman}, \&
  {Kocsis}}]{2020ApJ...898...25T}
{Tagawa}, H., {Haiman}, Z., \& {Kocsis}, B. 2020{\natexlab{b}}, \apj, 898, 25,
  \dodoi{10.3847/1538-4357/ab9b8c}

\bibitem[{{Tagawa} {et~al.}(2020{\natexlab{c}}){Tagawa}, {Kocsis}, {Haiman},
  {Bartos}, {Omukai}, \& {Samsing}}]{2020arXiv201200011T}
{Tagawa}, H., {Kocsis}, B., {Haiman}, Z., {et~al.} 2020{\natexlab{c}}, arXiv
  e-prints, arXiv:2012.00011.
\newblock \doarXiv{2012.00011}

\bibitem[{{Tagawa} {et~al.}(2020{\natexlab{d}}){Tagawa}, {Kocsis}, {Haiman},
  {Bartos}, {Omukai}, \& {Samsing}}]{2020arXiv201010526T}
---. 2020{\natexlab{d}}, arXiv e-prints, arXiv:2010.10526.
\newblock \doarXiv{2010.10526}

\bibitem[{{Takahashi} {et~al.}(2016){Takahashi}, {Ohsuga}, {Kawashima}, \&
  {Sekiguchi}}]{TOK16a}
{Takahashi}, H.~R., {Ohsuga}, K., {Kawashima}, T., \& {Sekiguchi}, Y. 2016,
  \apj, 826, 23, \dodoi{10.3847/0004-637X/826/1/23}

\bibitem[{{Takeo} {et~al.}(2018){Takeo}, {Inayoshi}, {Ohsuga}, {Takahashi}, \&
  {Mineshige}}]{2018MNRAS.476..673T}
{Takeo}, E., {Inayoshi}, K., {Ohsuga}, K., {Takahashi}, H.~R., \& {Mineshige},
  S. 2018, \mnras, 476, 673, \dodoi{10.1093/mnras/sty264}

\bibitem[{Tamborra \& Ando(2016)}]{Tamborra:2015fzv}
Tamborra, I., \& Ando, S. 2016, Phys. Rev., D93, 053010,
  \dodoi{10.1103/PhysRevD.93.053010}

\bibitem[{{Tanigawa} {et~al.}(2012){Tanigawa}, {Ohtsuki}, \&
  {Machida}}]{2012ApJ...747...47T}
{Tanigawa}, T., {Ohtsuki}, K., \& {Machida}, M.~N. 2012, \apj, 747, 47,
  \dodoi{10.1088/0004-637X/747/1/47}

\bibitem[{{Tanigawa} \& {Tanaka}(2016)}]{2016ApJ...823...48T}
{Tanigawa}, T., \& {Tanaka}, H. 2016, \apj, 823, 48,
  \dodoi{10.3847/0004-637X/823/1/48}

\bibitem[{{Tanigawa} \& {Watanabe}(2002)}]{2002ApJ...580..506T}
{Tanigawa}, T., \& {Watanabe}, S.-i. 2002, \apj, 580, 506,
  \dodoi{10.1086/343069}

\bibitem[{{Tanikawa} {et~al.}(2020){Tanikawa}, {Kinugawa}, {Yoshida},
  {Hijikawa}, \& {Umeda}}]{2020arXiv201007616T}
{Tanikawa}, A., {Kinugawa}, T., {Yoshida}, T., {Hijikawa}, K., \& {Umeda}, H.
  2020, arXiv e-prints, arXiv:2010.07616.
\newblock \doarXiv{2010.07616}

\bibitem[{{Thompson} {et~al.}(2005){Thompson}, {Quataert}, \&
  {Murray}}]{2005ApJ...630..167T}
{Thompson}, T.~A., {Quataert}, E., \& {Murray}, N. 2005, \apj, 630, 167,
  \dodoi{10.1086/431923}

\bibitem[{{Toomre}(1964)}]{Too64a}
{Toomre}, A. 1964, \apj, 139, 1217, \dodoi{10.1086/147861}

\bibitem[{Ueda {et~al.}(2014)Ueda, Akiyama, Hasinger, Miyaji, \&
  Watson}]{Ueda:2014tma}
Ueda, Y., Akiyama, M., Hasinger, G., Miyaji, T., \& Watson, M.~G. 2014,
  Astrophys.J., 786, 104, \dodoi{10.1088/0004-637X/786/2/104}

\bibitem[{{Vink} {et~al.}(2020){Vink}, {Higgins}, {Sander}, \&
  {Sabhahit}}]{2020arXiv201011730V}
{Vink}, J.~S., {Higgins}, E.~R., {Sander}, A. A.~C., \& {Sabhahit}, G.~N. 2020,
  arXiv e-prints, arXiv:2010.11730.
\newblock \doarXiv{2010.11730}

\bibitem[{{Waxman} \& {Katz}(2017)}]{2017hsn..book..967W}
{Waxman}, E., \& {Katz}, B. 2017, {Shock Breakout Theory}, ed. A.~W. {Alsabti}
  \& P.~{Murdin}, 967, \dodoi{10.1007/978-3-319-21846-5_33}

\bibitem[{{Weaver} {et~al.}(1977){Weaver}, {McCray}, {Castor}, {Shapiro}, \&
  {Moore}}]{1977ApJ...218..377W}
{Weaver}, R., {McCray}, R., {Castor}, J., {Shapiro}, P., \& {Moore}, R. 1977,
  \apj, 218, 377, \dodoi{10.1086/155692}

\bibitem[{{Wei} {et~al.}(2016){Wei}, {Cordier}, {Antier}, {Antilogus},
  {Atteia}, {Bajat}, {Basa}, {Beckmann}, {Bernardini}, {Boissier}, {Bouchet},
  {Burwitz}, {Claret}, {Dai}, {Daigne}, {Deng}, {Dornic}, {Feng}, {Foglizzo},
  {Gao}, {Gehrels}, {Godet}, {Goldwurm}, {Gonzalez}, {Gosset}, {G{\"o}tz},
  {Gouiffes}, {Grise}, {Gros}, {Guilet}, {Han}, {Huang}, {Huang}, {Jouret},
  {Klotz}, {La Marle}, {Lachaud}, {Le Floch}, {Lee}, {Leroy}, {Li}, {Li}, {Li},
  {Liang}, {Lyu}, {Mercier}, {Migliori}, {Mochkovitch}, {O'Brien}, {Osborne},
  {Paul}, {Perinati}, {Petitjean}, {Piron}, {Qiu}, {Rau}, {Rodriguez},
  {Schanne}, {Tanvir}, {Vangioni}, {Vergani}, {Wang}, {Wang}, {Wang}, {Wang},
  {Watson}, {Webb}, {Wei}, {Willingale}, {Wu}, {Wu}, {Xin}, {Xu}, {Yu}, {Yu},
  {Yu}, {Zhang}, {Zhang}, {Zhang}, \& {Zhou}}]{2016arXiv161006892W}
{Wei}, J., {Cordier}, B., {Antier}, S., {et~al.} 2016, arXiv e-prints,
  arXiv:1610.06892.
\newblock \doarXiv{1610.06892}

\bibitem[{{Woosley}(2017)}]{2017ApJ...836..244W}
{Woosley}, S.~E. 2017, \apj, 836, 244, \dodoi{10.3847/1538-4357/836/2/244}

\bibitem[{{Yang} {et~al.}(2020){Yang}, {Gayathri}, {Bartos}, {Haiman},
  {Safarzadeh}, \& {Tagawa}}]{2020ApJ...901L..34Y}
{Yang}, Y., {Gayathri}, V., {Bartos}, I., {et~al.} 2020, \apjl, 901, L34,
  \dodoi{10.3847/2041-8213/abb940}

\bibitem[{{Yang} {et~al.}(2019){Yang}, {Bartos}, {Gayathri}, {Ford}, {Haiman},
  {Klimenko}, {Kocsis}, {M{\'a}rka}, {M{\'a}rka}, {McKernan}, \&
  {O'Shaughnessy}}]{2019PhRvL.123r1101Y}
{Yang}, Y., {Bartos}, I., {Gayathri}, V., {et~al.} 2019, \prl, 123, 181101,
  \dodoi{10.1103/PhysRevLett.123.181101}

\bibitem[{Yuan {et~al.}(2020)Yuan, Murase, Kimura, \&
  M\'esz\'aros}]{Yuan:2020oqg}
Yuan, C., Murase, K., Kimura, S.~S., \& M\'esz\'aros, P. 2020, Phys. Rev. D,
  102, 083013, \dodoi{10.1103/PhysRevD.102.083013}

\bibitem[{Yuan {et~al.}(2021)Yuan, Murase, Zhang, Kimura, \&
  M\'esz\'aros}]{Yuan:2021jjt}
Yuan, C., Murase, K., Zhang, B.~T., Kimura, S.~S., \& M\'esz\'aros, P. 2021.
\newblock \doarXiv{2101.05788}

\bibitem[{{Zhu} {et~al.}(2020){Zhu}, {Zhang}, {Yu}, \&
  {Gao}}]{2020arXiv201108428Z}
{Zhu}, J.-P., {Zhang}, B., {Yu}, Y.-W., \& {Gao}, H. 2020, arXiv e-prints,
  arXiv:2011.08428.
\newblock \doarXiv{2011.08428}

\end{thebibliography}

\listofchanges

\end{document}